%% file: main.tex
\documentclass[a4paper,11pt]{article}
\pdfoutput=1 % if your are submitting a pdflatex (i.e. if you have
\usepackage{jcappub} % for details on the use of the package, please
\usepackage[T1]{fontenc} % if needed

\input{preamble}

\input{Befehle}

\usepackage{bbm}

\title{Evolution of neutrino distributions under repeated electron scattering}

\author[a]{Jens Chluba}
\author[b]{, Bryce Cyr}
\author[c]{, Michiru Uwabo-Niibo}
\author[c,d,e]{and Masahide Yamaguchi}

\affiliation[a]{Jodrell Bank Centre for Astrophysics, School of Physics and Astronomy, The University of Manchester, Oxford Road, Manchester, M13 9PL, U.K.}
\affiliation[b]{Center for Theoretical Physics - A Leinweber Institute, Massachusetts Institute of Technology, Cambridge, MA 02139, USA}
\affiliation[c]{Cosmology, Gravity, and Astroparticle Physics Group, Center for
Theoretical Physics of the Universe, Institute for Basic Science (IBS), Daejeon}
\affiliation[d]{Department of Physics, Institute of Science Tokyo, 2-12-1 Ookayama, Meguro-ku, Tokyo 152-8551, Japan}
\affiliation[e]{Department of Physics and IPAP, Yonsei University, 50 Yonsei-ro, Seodaemun-gu, Seoul 03722, Korea}

\date{September 2026}
\begin{document}
\begin{flushright}
  \texttt{MIT-CTP/6125}
\end{flushright}
\vspace{-2em}

\abstract{We study the energy redistribution of neutrinos by repeated elastic scattering with an isotropic thermal bath of free electrons. Starting from the relevant Boltzmann collision integral, we formulate the evolution equations in terms of the neutrino--electron redistribution kernel derived in our companion work. The scattering kernel formulation retains recoil, relativistic kinematics and Pauli blocking, while reducing the calculation to a one-dimensional integral in the outgoing neutrino energy.  To gain a better analytic understanding, we derive the corresponding Fokker--Planck (FP) limit of the collision operator in two independent ways and contrast it with the Kompaneets equation for photons. This shows that the different energy dependence of the weak cross section changes the drift and diffusion operators which causes the high-energy part of a neutrino distribution to evolve much faster than its low-energy part. We discuss the fermionic analogues of temperature-, chemical-potential- and $y$-type spectral distortions, and give approximate analytic solutions to the FP equation in several limiting cases.
We also derive the analytic expressions for the relativistic scattering operator when Doppler terms dominate, allowing us to compute the relativistic Sunyaev-Zeldovich equivalent for neutrinos in the inefficient scattering limit. We compare these expressions to the full numerical result based on the detailed collision term.
Finally, we set up conservative numerical schemes for both the FP equation and the scattering kernel, generalizing previous codes developed for studying the repeated Compton scattering of photons by relativistic electrons. Controlled line-injection experiments then illustrate when the diffusion approximation is accurate and where the non-locality of the redistribution kernel remains essential.
The results of this study provide useful analytic and numerical benchmarks for energy-redistribution problems relevant to early-Universe applications and supernova neutrino transport problems.}

\maketitle

\section{Introduction}
%---------------------------------------
Elastic scattering of neutrinos by free electrons is one of the basic processes that redistributes neutrino energies in the early Universe \citep[e.g.,][]{Dolgov2002, Hannestad1995, Bond2024nu} and in hot and dense astrophysical environments such as supernovae \citep[e.g.,][]{Bruenn1985,Burrows1995, Thompson2000, Burrows2002,Liebendorfer2005,Buras2006}. 
The microscopic ingredients for describing this process are well known, but the evolution generated by many scatterings is less studied: the weak cross section is strongly energy dependent, recoil competes with Doppler broadening, and at high densities Pauli blocking can affect both the target electron and neutrino final states. A direct Boltzmann calculation contains all of this physics, but hides the redistribution associated with individual scattering events inside a multi-dimensional phase-space integral.

In understanding some of these details, the closely related problem of repeated Compton scattering provides a useful guide. For photons, the interplay between Doppler broadening, recoil and stimulated scattering can be described in a variety of complementary ways, ranging from the Kompaneets equation and analytic limiting solutions to exact redistribution-kernel methods \citep[e.g.,][]{Kompa56, Pozdniakov1979, Nagirner1994, Belmont2008, CSpack2019}. These approaches have been particularly useful for understanding the evolution of spectral distortions of the cosmic microwave background (CMB) and the approach towards kinetic equilibrium \citep{Illarionov1975, Burigana1991, Hu1993, Chluba2011therm}. Neutrino scattering shares many elements of this structure, while differences in the microscopic interaction and fermionic statistics lead to distinct redistribution dynamics. A systematic comparison therefore provides a convenient way of identifying which aspects of thermal scattering are generic and which are specific to the underlying interaction.

In a companion paper \citep{Chluba2026nuKernels} we derived the isotropic neutrino--electron redistribution kernel, retaining the full scattering kinematics and reducing the collision integral to a form suitable for efficient numerical evaluation. Here we use this kernel as an evolution operator and investigate the physics of repeated energy redistribution. We derive the corresponding Fokker--Planck description in two independent ways, using energy-transfer moments and the boost-operator formalism, and study several limiting cases analytically. These solutions allow us to isolate the roles of Doppler broadening, recoil and background-induced scattering and to directly contrast neutrino and photon evolution. The same framework can also be mapped onto other elastic neutrino scattering processes, as we show here.

We furthermore use the boost-operator approach \citep{ChlubaBO25, Rosenberg2025prSZ, Hoey2026Kompaneets} to derive the relativistic single-scattering operator for thermal electrons and construct an economic spectral basis for the resulting neutrino background distortion. This provides a neutrino analogue of the relativistic Sunyaev--Zeldovich effect \citep{Zeldovich1969, Sunyaev1980ARAA} and, more generally, a controlled example in which relativistic redistribution effects can be studied analytically. We then develop numerical schemes for both the Fokker--Planck equation and the full redistribution kernel, generalising methods previously applied to repeated Compton scattering \citep{Acharya2021, Acharya2022large}. Controlled line-injection and small-distortion experiments allow us to follow the transition from individual scattering events towards kinetic equilibrium and to identify where local diffusion approximations cease to capture the full redistribution dynamics.
The primary aim of this work is therefore not to model one particular observable or astrophysical environment, but to develop physical intuition, analytic benchmarks and numerical tools for neutrino energy-redistribution problems. The comparison with photon Comptonization is useful in this regard because it places the neutrino problem within a familiar kinetic framework while exposing the consequences of the different interaction physics. The methods developed here can subsequently be incorporated into more detailed cosmological or astrophysical transport calculations.

The paper is organised as follows. In section~\ref{sec:exact_problem} we formulate the neutrino--electron collision term in terms of the redistribution kernel. In section~\ref{sec:FP}, we derive the Fokker--Planck limit of the collision term using both energy-transfer moments and the boost-operator approach, and extend the result to other elastic neutrino scattering processes. Section~\ref{sec:analytics} develops analytic solutions in several limiting regimes and compares them with their photon counterparts. In section~\ref{sec:nuSZ_effect} we derive the relativistic single-scattering, or $\nu$SZ, operator and introduce a spectral basis for describing the signal. Section~\ref{sec:sing_scat_col} evaluates the exact collision term for thermal distributions and illustrates the relativistic single-scattering limit. In section~\ref{sec:num_setup} we describe the numerical treatment of both the Fokker--Planck and redistribution-kernel problems and discuss the main spectral shapes close to equilibrium. Section~\ref{sec:num_sols} presents numerical examples of repeated scattering, including line injection and the relaxation of spectral distortions towards kinetic equilibrium. As an example, in section~\ref{sec:y_sc} we estimate the efficiency of neutrino--electron scattering and energy redistribution around the epoch of electron--positron annihilation, before concluding in section~\ref{sec:conclusions}.

\section{Kinetic equation in the kernel formulation}
\label{sec:exact_problem}
%---------------------------------------
We consider the elastic process $\nu(P_1)+e(P_2)\leftrightarrow\nu(P_3)+e(P_4)$ and assume isotropic
fermionic distribution functions, $f_i(t, E_i)$, for the involved particles $i=\{1, 2, 3, 4\}$ with four vectors $P_i$ and energy $E_i$. In a homogeneous expanding medium the relevant Boltzmann equation is then given by
%-------------------------------------
\begin{align}
\label{eq:boltzmann}
\left(\partial_t-Hp_1\partial_{p_1}\right)f_1=\mathcal{C}[f_1],
\end{align}
%-------------------------------------
where $p_i=q_i/\me$ is the physical momentum in rest mass units ($q_i$ being the usual momentum variable),  $H=\dot{a}/a$ denotes the Hubble factor and $a$ is the scale-factor in the problem. Using the dimensionless neutrino energy, $x=E/T_{\rm r}$, with a reference temperature satisfying $T_{\rm r}\propto a^{-1}$, one can absorb the expansion term when changing variables from $f(t, E)\rightarrow f(t, x)$, yielding
%-------------------------------------
\begin{align}
\label{eq:boltzmann_fx}
\partial_t f_1=\mathcal{C}[f_1].
\end{align}
%-------------------------------------
This means that, similar to the scattering of photons by free electrons in the expanding Universe \citep[e.g.,][]{Burigana1991, Hu1993, Chluba2011therm}, one can study the evolution of neutrino distributions in both static and expanding media using the same evolution equation.
Restoring the full phase-space expression for reference, the relevant collision term is \citep[e.g.,][]{Hannestad1995, Burrows2002, Bond2024nu,Chluba2026nuKernels}
%-------------------------------------
\bsub
\label{eq:collision_full}
\begin{align}
\mathcal{C}[f_1]={}&\int\frac{\id^3q_2}{(2\pi)^3}
\frac{\id^3q_3}{(2\pi)^3}\frac{\id^3q_4}{(2\pi)^3}
\frac{(2\pi)^4\delta^{(4)}(P_4+P_3-P_2-P_1)}{2^4E_1E_2E_3E_4}
\,|\mathcal{M}|^2\,\mathcal{F}_{1234},
\\[3mm]
\mathcal{F}_{1234}={}&f_3f_4(1-f_1)(1-f_2)
-f_1f_2(1-f_3)(1-f_4).
\end{align}
\esub
%-------------------------------------
Here,  $|\mathcal{M}|^2$ is the matrix element for the relevant scattering process (summed over initial and final spin-states). The distribution functions $f_1$ and $f_3$ respectively refer to the incoming and scattered neutrino, while $f_2$ and $f_4$ refer to the corresponding electrons.  The first and second terms in the statistical factor $\mathcal{F}_{1234}$ describe gain and loss, respectively.  Equation~\eqref{eq:collision_full} makes explicit that Pauli blocking [i.e., the $(1-f_i)$ factors] enters for both final-state particles. For illustration, we will consider examples with degenerate and non-degenerate distribution functions for the electrons and neutrinos.

After using energy--momentum conservation and carrying out the angular integrations, the redistribution kernel version of Eq.~\eqref{eq:collision_full} becomes \citep{Chluba2026nuKernels}
%-------------------------------------
\begin{align}
\label{eq:collision_kernel_general}
\mathcal{C}[f_1]=\int_0^\infty\frac{q_2^2\id q_2}{2\pi^2}
\int_0^\infty\id\omega_3\,
P(\omega_1\rightarrow\omega_3,p_2)\,\mathcal{F}_{1234},
\end{align}
%-------------------------------------
with kernel $P(\omega_1\rightarrow\omega_3,p_2)$ for $p_2=q_2/\me$, $\omega_i=E_i/\me$ for the neutrinos, $\gamma_i=E_i/\me=\sqrt{1+p^2_i}$ for the electrons, and energy of the scattered electron $\gamma_4=\omega_1+\gamma_2-\omega_3$. This form for the collision term is a natural starting point because all distribution functions depend only on particle energies.

The kernel $P(\omega_1\rightarrow\omega_3,p_2)$ contains the kinematic factors and the weak matrix element; its explicit analytic form and integration zones are given in \citep{Chluba2026nuKernels}. One can furthermore use a similar expression for the collision term of the Compton process \citep{Chluba2020large, Acharya2022large} and other neutrino scattering channels once switching to the appropriate redistribution kernels and statistical factors. Efficient numerical discretisation schemes will be discussed below and follow from previous treatments of the photon scattering problem \citep{Chluba2020large, Acharya2022large} as now implemented in {\tt CSpack} \citep{CSpack2019}.

\subsection{Thermal kernel and the collision term}
%-------------------------------------
To understand how neutrinos scatter with thermal electrons more generally, we consider the collision term assuming equilibrium distributions at varying temperatures, $\theta=T/\me$. If the neutrino and electron temperatures are equal, $\theta_\nu=\The$, the collision term vanishes for constant chemical potentials. This can be easily seen when writing the statistical factor explicitly
%-------------------------------------
\begin{align}
\label{eq:F_thermal}
\mathcal{F}_{1234}&=f_3 f_4 (1-f_1) (1-f_2)-f_1 f_2 (1-f_3) (1-f_4)
\nonumber\\
&=f_2(1-f_4)\times f_1 (1-f_3) 
\times\left\{\frac{1-f_2}{f_2}\frac{f_4}{1-f_4}\times\frac{1-f_1}{f_1}\frac{f_3}{1-f_3}  -1 \right\}
\nonumber\\
&=\frac{\expf{(\mu_2-\gamma_2)/\The}}{[1+\expf{(\mu_2-\gamma_2)/\The}][1+\expf{(\mu_4-\gamma_4)/\The}]}\times \frac{\expf{(\mu_1-\omega_1)/\theta_\nu}\left\{\expf{(\gamma_2-\gamma_4)/\The}\times\expf{(\omega_1-\omega_3+\mu_3-\mu_1)/\theta_\nu}  -1 \right\}}{[1+\expf{(\mu_1-\omega_1)/\theta_\nu}][1+\expf{(\mu_3-\omega_3)/\theta_\nu}]}
\nonumber\\
&=\frac{\expf{-z_2}}{[1+\expf{-z_2}][1+\expf{-z_4}]}\times \frac{\expf{\hat{\mu}_1-x_1}\, 
\left\{\expf{(x_3-x_1)(\phi-1)}\times\expf{\hat{\mu}_3-\hat{\mu}_1}  -1 \right\}}{[1+\expf{\hat{\mu}_1-x_1}][1+\expf{\hat{\mu}_3-x_3}]},
\end{align}
%-------------------------------------
where we used $(1-f_i)/f_i=\expf{(\gamma_i-\mu_i)/\theta_i}$ and $\gamma_2-\gamma_4=\omega_3-\omega_1$, while defining $z_i=(\gamma_i-\mu_i)/\The$, $x_i=\omega_i/\theta_\nu$, $\hat{\mu}_i=\mu_i/\theta_\nu$ and $\phi=\theta_\nu/\The$. For the electron distributions we directly assumed $\mu_2=\mu_4$ (i.e., equilibrium Fermi-Dirac distributions with energy-independent chemical potential), while for the neutrino distributions we left matters general with $\mu_1\neq \mu_3$. 
Similarly, we can write
%-------------------------------------
\begin{align}
\label{eq:F_thermal_photon}
\mathcal{F}^\gamma_{1234}&=f_3 f_4 (1+f_1) (1-f_2)-f_1 f_2 (1+f_3) (1-f_4)
\nonumber\\
&=\frac{\expf{-z_2}}{[1+\expf{-z_2}][1+\expf{-z_4}]}\times \frac{\expf{\hat{\mu}_1-x_1}\, 
\left\{\expf{(x_3-x_1)(\phi-1)}\times\expf{\hat{\mu}_3-\hat{\mu}_1}  -1 \right\}}{[1-\expf{\hat{\mu}_1-x_1}][1-\expf{\hat{\mu}_3-x_3}]},
\end{align}
%-------------------------------------
for photon-electron/positron scattering processes, where it is understood that $\phi$, $x_i$ and $\hat{\mu}_i$ are now all evaluated with $\Thg$ instead of $\theta_\nu$. From these two versions of the statistical factor we can directly see that if $\hat{\mu}_1=\hat{\mu}_3$ (i.e., chemical equilibrium under the scattering reaction) and $\phi=1$ the collision term vanishes. Departures from either of these cases lead to net energy exchange and modifications of the neutrino distribution functions under scattering.

We next define the thermally-averaged scattering kernel per electron as
%-------------------------------------
\begin{align}
\label{eq:P_therm}
P(\omega_1 \rightarrow \omega_3,\The, \mu_{\rm e})&=
\frac{1}{\Ne}\,\int 
\frac{q_2^2\id q_2}{2\pi^2}  \frac{\expf{-z_2} P(\omega_1 \rightarrow \omega_3, p_2)}{[1+\expf{-z_2}][1+\expf{-z_4}]},
\qquad \Ne=\int 
\frac{q_2^2\id q_2}{\pi^2}  \frac{\expf{-z_2}}{1+\expf{-z_2}},
\end{align}
%-------------------------------------
where we divide by the total electron number density, $\Ne$, as required. Due to energy conservation, we can use $\gamma_4=\omega_1+\gamma_2-\omega_3$, which means this averaged kernel can be pre-computed as a function of $\omega_1$ and $\omega_3$ given $\The$ and $\mu_{\rm e}$. This naturally defines a scattering matrix for the energy redistribution problem as we explain below.
The same definition applies to all scattering processes involving photons, electrons or positrons where the scattering kernel simply has to be exchanged appropriately.
The collision term per electron can then be compactly written as
%-------------------------------------
\begin{align}
\label{eq:C_def_kernel_eval}
\mathcal{C}_{\pm}
&=
\int \id \omega_3 \, P(\omega_1 \rightarrow \omega_3,\The, \mu_{\rm e})\times 
\frac{\expf{\hat{\mu}_1-x_1}\, 
\left\{\expf{(x_3-x_1)(\phi-1)}\times\expf{\hat{\mu}_3-\hat{\mu}_1}  -1 \right\}}{[1\pm \expf{\hat{\mu}_1-x_1}][1\pm\expf{\hat{\mu}_3-x_3}]},
\end{align}
%-------------------------------------
where '$+$' is for neutrino scattering events and '$-$' for photon scattering. This is still the general version of the collision term in the presence of thermal electrons/positrons. Since the thermally-averaged kernel can be pre-computed, this version provides the starting point for efficient numerical evaluations as we will explain below \citep[e.g.,][for photon scattering]{Chluba2020large, Acharya2021}.

%\subsection{Detailed balance relation}
\paragraph{Detailed balance relation.}
%-------------------------------------
For completeness we note that for a thermal target bath of electron the forward and reverse kernels obey the detailed balance relation
%-------------------------------------
\begin{align}
\label{eq:kernel_detailed_balance}
P(\omega_3\rightarrow\omega_1,\The,\mu_{\rm e})
=\frac{\omega_1^2}{\omega_3^2}
\expf{\frac{\omega_3-\omega_1}{\The}}
P(\omega_1\rightarrow\omega_3,\The,\mu_{\rm e}).
\end{align}
%-------------------------------------
This identity provides both a stringent numerical check and the simplest proof
that a Fermi--Dirac spectrum at the electron temperature, with an arbitrary
constant neutrino chemical potential, is stationary under elastic scattering. The same detailed balance relation applies to photon-electron scattering kernels \citep{Sazonov2000, CSpack2019}.

\section{Fokker--Planck approximation for the neutrino collision integral}
\label{sec:FP}
%-------------------------------------------------
We next derive the Fokker--Planck (FP) approximations for the collision term. This limit is useful both as an efficient evolution scheme and as a guide to the
physics of drift and broadening in the diffusive regime.  We give two derivations: a direct expansion in the energy transfer, followed by a derivation based on the boost operator approach \citep[see][]{ChlubaBO25, Hoey2026Kompaneets}. The latter also allows us to obtain an all orders included expression for scattering operator in the Doppler-dominated regime, as we explain in section~\ref{sec:nuSZ_effect}.

\subsection{Fokker--Planck approximation using the energy transfer moments up to second order}
%-------------------------------------------------
To obtain the FP approximation for the collision integral, we first rewrite the statistical factor as 
%-------------------------------------
\begin{align}
\label{eq:F_coll}
\mathcal{F}_{1234} &=
        f_{3}f_{4}(1-f_{1})(1-f_{2})
        -
        f_{1}f_{2}(1-f_{3})(1-f_{4})=f_{2}(1-f_{4})\left[\frac{f_4(1-f_{2})}{f_2(1-f_4)}f_{3}(1-f_1)-f_1(1-f_3)\right]
\nonumber\\
&=f_{2}(1-f_{4})\left[\expf{\Delta x}\,f_{3}(1-f_1)-f_1(1-f_3)\right]
\end{align}
%-------------------------------------
where we used that $f_4(1-f_{2})/[f_2(1-f_4)]=\exp[(\omega_3-\omega_1)/\The]$ and introduced $\Delta x=(\omega_3-\omega_1)/\The$. We are interested in the diffusion limit of the problem, which means we have to consider the energy shift at second order in $\Delta=\omega_3/\omega_1-1$. We can then use $$f_3\approx f_1+\Delta\,\omega_1 \partial_{\omega_1} f_1 +\frac{\Delta^2}{2} \omega_1^2 \partial^2_{\omega_1} f_1 
=f-\Delta \,\boostO f  +\frac{\Delta^2 }{2}\boostO(\boostO+1) f $$ in the statistical factor, where $\boostO=-\omega_1 \partial_{\omega_1}$ and we wrote $f=f_1$ for convenience. Inserting this back into Eq.~\eqref{eq:F_coll}, we then have
%-------------------------------------
\begin{align}
\mathcal{F}_{1234} &\approx 
f_{2}(1-f_{4})\left[\Delta \{x f(1-f)-\boostO f\}+\frac{\Delta^2}{2}
\{
\boostO(\boostO+1) f
-2 x (1-f) \boostO f + x^2 f(1-f)
\}\right]
\end{align}
%-------------------------------------
Through the term $f_2 (1-f_4)$,  this expression still accounts for the effects of Fermi-blocking in the electrons. However, at leading order in the electron temperature, determined by the terms in the bracket alone, we can set $f_{2}(1-f_{4})\approx f_{2}$.
Inserting this into the collision integral, using the thermally-averaged moments \citep{Chluba2026nuKernels}, $\Sigma_1=\langle \Delta\rangle\approx \The^3 \bar{\sigma}_1 x^2 \lambda_{LR} (6-x)$ and $\Sigma_2=\langle \Delta^2\rangle\approx 2 \The^3 \bar{\sigma}_1 x^2 \lambda_{LR}$ (with $x=\omega_1/\The$), and collecting terms, the collision term in Eq.~\eqref{eq:C_def_kernel_eval} becomes
%-------------------------------------
\begin{align}
\label{eq:FP}
\mathcal{C}_+ &
\approx 
\Sigma_1 \{x f(1-f)-\boostO f\}+\frac{\Sigma_2}{2}
\{
\boostO(\boostO+1) f
-2 x (1-f) \boostO f + x^2 f(1-f)
\}\nonumber\\
&=
\The^3 \bar{\sigma}_1 x^2 \, \lambda_{LR}\left[(6-x) \{xf(1-f)-\boostO f\}+
\boostO(\boostO+1) f
-2 x (1-f) \boostO f + x^2 f(1-f)\right]
\nonumber\\
&=
\The^3 \bar{\sigma}_1 x^2 \, \lambda_{LR}\left[(6-x) \{x f(1-f)-\boostO f\}
+\boostO(\boostO-5) f
+6\boostO f
-2 x (1-f) \boostO f + x^2 f(1-f)\right]
\nonumber\\
&=
\The^3 \bar{\sigma}_1 x^2 \, \lambda_{LR}\left[
\boostO(\boostO-5) f
+ x \boostO f+ 6 x f(1-f)
-2 x \boostO f + 2 x f \boostO f \right]
\nonumber\\
&=
\The^3 \bar{\sigma}_1 x^2 \, \lambda_{LR}\left[
\boostO(\boostO-5) f
+ 6 x f(1-f)
- x (1-2  f) \boostO f \right]
\nonumber\\
&=
\The^3 \bar{\sigma}_1 x^2 \, \lambda_{LR}\left[
\boostO(\boostO-5) f
+ x (6 - \boostO) f(1-f)\right]
\nonumber\\
&=\The^3 \bar{\sigma}_1 \lambda_{LR} \frac{1}{x^2}\partial_x x^6 [\partial_x f + f (1-f)],
\end{align}
%-------------------------------------
where we used that $x^2 \boostO(\boostO-5) = \frac{1}{x^2}\partial_x x^6 \partial_x$ and $x^3 (6 - \boostO) = \frac{1}{x^2}\partial_x x^6$. In addition, the leading cross section averaged over initial electron spins is $\bar{\sigma}_1\approx \pot{1.11}{-20}\sigT$ at $\omega_1=1$ and $\lambda_{LR}\approx 0.87$ for $\nu_e-e$-scattering \citep[see][for more details]{Chluba2026nuKernels}. This FP approximation has a similar form as the result of \citep{Suwa2019nuKomp} obtained for neutrino-nuclear scattering when neglecting bulk motions. We will refer to Eq.~\eqref{eq:FP} as the \emph{neutrino Kompaneets equation} for $\nu_e-e$-scattering in analogy to the photon Kompaneets equation \citep{Kompa56}.  The corresponding low-temperature expansion for photon contains $x^{-2}\partial_x x^4[\partial_x f+f(1+f)]$: the factor $x^6$ for neutrinos is replaced by $x^4$ for photons and Fermi blocking by Bose stimulation $(1-f)\rightarrow (1+f)$. The extra factor $x^2$ in the neutrino operator directly reflects the low-energy cross section scaling $\sigma_{\nu e}\propto E_\nu^2$ and causes the high energy part of the spectrum to evolve more quickly.

%-------------------------------------------------
\subsection{Fokker--Planck approximation using boost operator approach}
\label{sec:FP_nu}
%-------------------------------------------------
We can also derive the FP approximation of the problem using the boost operator approach \citep{ChlubaBO25}. This was recently demonstrated for the photon-electron scattering problem \citep{Hoey2026Kompaneets} and here can be followed very similarly. We start by writing the collision term in the electron frame 
%---------------------------
\begin{align}
\label{app:col_term_int}
\frac{\text{d}f(\omega_1, \vgh_1)}{\text{d}\tau}  
&= \int \frac{\text{d}\vgh_3}{4\pi}\,
\Bigg\{\frac{\tilde{\omega}^4_3}{\omega^4_1}\frac{\text{d}\sigma(\tilde{\omega}_3,\omega_1,\chi_{13})}{\text{d}\Omega}\,f(\tilde{\omega}_3, \vgh_3)[1-f(\omega_1, \vgh_1)]
\nonumber \\
&\qquad \qquad \qquad \qquad - \frac{\text{d}\sigma(\omega_1, \omega_3, \chi_{13})}{\text{d}\Omega}\,f(\omega_1, \vgh_1)[1-f(\omega_3, \vgh_3)]\Bigg\},
\end{align}
%---------------------------
where $\omega_3=\omega_1/[1+\omega_1(1-\chi_{13})]$ is the scattered neutrino energy with $\chi_{13}=\vgh_1\cdot \vgh_3$.
We also have 
$\tilde{\omega}_3=\omega_1/[1-\omega_1(1-\chi_{13})]$, as required to ensure that the scattered neutrino has energy $\omega_1$ when running the process backwards for the first term in the integral. The factor $\tilde{\omega}^4_3/\omega^4_1=(\tilde{\omega}_3/\omega_1)^2 \id \tilde{\omega}_3/\id \omega_1$ in the gain term comes from the transformation of the neutrino phase space elements.\footnote{For the Compton scattering case $\frac{\tilde{\omega}^4_3}{\tilde{\omega}^4_1}\frac{\text{d}\sigma(\tilde{\omega}_3,\omega_1,\chi_{13})}{\text{d}\Omega}\equiv \frac{\text{d}\sigma(\omega_1,\tilde{\omega}_3, \chi_{13})}{\text{d}\Omega}$ as was used by \citep{Hoey2026Kompaneets}.} 
Fermi-blocking was directly neglected for the electrons, as the related effects will enter at higher order in $\The$ as before.

The arguments go as follows: we are interested in terms up to order $\omega_1$ to account for the leading order electron recoil effect. We can therefore write
%---------------------------
\bealf{
\label{app:photon_occ_expansion_omega_K}
f(\tilde{\omega}_3,\vgh_3) &\approx f(\omega_1,\vgh_3) - \omega_1\,(1-\chi_{13})\,\hat{\mathcal{O}}_{\omega_1}f(\omega_1), 
\nonumber\\
f(\omega_3,\vgh_3) &\approx f(\omega_1,\vgh_3) + \omega_1\,(1-\chi_{13})\,\hat{\mathcal{O}}_{\omega_1}f(\omega_1),
}
%---------------------------
where for the terms $\propto \omega_1$ we can immediately use the isotropic part of the distribution, as other contributions would be of higher order in the electron temperature once transformed back in the lab frame. For the rest frame differential cross section, we have \citep{Chluba2026nuKernels}
%-------------------------------------
\begin{align}
\frac{\id \sigma^{\rm rest}(\omega_1, \omega_3)}{\id \Omega}
&=\bar{\sigma}_1 \omega_3^2\left\{1+\alpha_{LR}\left[1-\frac{\omega_3}{\omega_1}(1-\chi_{13})\right] - \beta_{LR} \omega_3 (1-\chi_{13})\,\left[2-\omega_3(1-\chi_{13})\right]\right\}
\end{align}
%-------------------------------------
with $\alpha_{LR}=g_{L}g_{R}/(g_{L}^{2}-g_{L}g_{R}+g_{R}^{2})=3(1-\lambda_{LR})$ and $\beta_{LR}=g_{R}^{2}/(g_{L}^{2}-g_{L}g_{R}+g_{R}^{2})$ arising from differences in the left and right-handed couplings. 
Keeping terms up to $\omega_1^3$ this then means 
%-------------------------------------
\begin{align}
\frac{\id \sigma^{\rm rest}(\omega_1, \omega_3)}{\id \Omega}
&\approx \bar{\sigma}_1 \omega_1^2
\Bigg\{1+\alpha_{LR}\, \chi_{13}-\omega_1\left[2-\alpha_{LR}+2\beta_{LR}+3\alpha_{LR}\,\chi_{13}\right](1-\chi_{13})
\Bigg\},
\nonumber\\
\frac{\tilde{\omega}^4_3}{\omega^4_1}\,\frac{\id \sigma^{\rm rest}(\tilde{\omega}_3, \omega_1)}{\id \Omega}
&\approx \bar{\sigma}_1 \omega_1^2
\Bigg\{1+\alpha_{LR}\, \chi_{13}+\omega_1\left[4+\alpha_{LR}-2\beta_{LR}+3\alpha_{LR}\,\chi_{13}\right](1-\chi_{13})
\Bigg\}.
\end{align}
%-------------------------------------
The overall cross section scaling is $\propto \bar{\sigma}_1 \omega_1^2$, while terms $\propto \omega_1^3$ capture the leading order electron recoil effect. Importantly, we will see that in the rest frame these terms simplify significantly once the integral over the scattering angle, $\theta_{13}=\arccos(\chi_{13})$, is taken.

Inserting this into Eq.~\eqref{app:col_term_int} and dropping terms that are of higher order in $\omega_1$, we can evaluate the integral over the scattering angle:
%---------------------------
\begin{align}
\label{app:col_term_int_approx}
\frac{\text{d}f(\omega_1, \vgh_1)}{\text{d}\tau}  
&\approx  \bar{\sigma}_1 \omega_1^2\, \int \frac{\text{d}\vgh_3}{4\pi}\,
\Big\{(1+\alpha_{LR}\,\chi_{13}) [f(\omega_1, \vgh_3)-f(\omega_1, \vgh_1)]
\\
&\qquad \qquad \qquad \quad\;
+\omega_1\left[4+\alpha_{LR}-2\beta_{LR}+3\alpha_{LR}\,\chi_{13}\right](1-\chi_{13})
\,f(\omega_1, \vgh_3)[1-f(\omega_1, \vgh_1)]
\nonumber \\[1mm]
&\qquad \qquad \qquad \qquad \; +\omega_1\left[2-\alpha_{LR}+2\beta_{LR}+3\alpha_{LR}\,\chi_{13}\right](1-\chi_{13})\,f(\omega_1, \vgh_1)[1-f(\omega_1, \vgh_3)]
\nonumber \\[1mm]\nonumber 
&\qquad \qquad \qquad \qquad  \quad
-\omega_1 (1+\alpha_{LR}\,\chi_{13}) (1-\chi_{13})\left\{[1-f(\omega_1)] \hat{\mathcal{O}}_{\omega_1} f(\omega_1)
- f(\omega_1) \, \hat{\mathcal{O}}_{\omega_1} f(\omega_1)\right\}
\Big\}
\\[-1mm] \nonumber 
& 
\approx  \bar{\sigma}_1 \omega_1^2\, \int \frac{\text{d}\vgh_3}{4\pi}\,
\Bigg\{
(1+\alpha_{LR}\,\chi_{13}) [f(\omega_1, \vgh_3)-f(\omega_1, \vgh_1)]
\\ \nonumber 
&\qquad  \qquad  \qquad \qquad 
+6\omega_1 \left(1+\alpha_{LR}\,\chi_{13}\right)(1-\chi_{13})\,f(\omega_1)[1-f(\omega_1)]
\\\nonumber 
&\qquad \qquad \qquad  \qquad \quad 
-\omega_1 (1+\alpha_{LR}\,\chi_{13}) (1-\chi_{13})[1-2f(\omega_1)] \hat{\mathcal{O}}_{\omega_1} f(\omega_1)
\Bigg\}
\\ \nonumber
&\approx
\bar{\sigma}_1 \omega_1^2
\left\{
f_0+\frac{\alpha_{LR}}{3} f_1-f+
6\omega_1\left(1-\frac{\alpha_{LR}}{3}\right)\,f_0[1-f_0]
-\omega_1 \left(1-\frac{\alpha_{LR}}{3}\right)[1-2 f_0] \hat{\mathcal{O}}_{\omega_1} f_0
\right\}
\end{align}
%---------------------------
where $f_\ell=f_\ell(\omega_1, \vgh)=\sum_{m} f_{\ell m}(\omega) Y_{ \ell m}(\vgh) $ and $f=f(\omega_1, \vgh)$. We note that $1-\frac{\alpha_{LR}}{3}=\lambda_{\rm LR}$. 
To transform back to the lab frame we have to consider the terms $f_0+\frac{1}{3}\alpha_{LR} f_1-f$ carefully, while the other terms just carry over. Because the cross section depends on $\omega_1^2$, this means that the Doppler operators \citep{ChlubaBO25, Hoey2026Kompaneets}
%---------------------------
\begin{align}
{}^{-3}\hat{\mathcal{D}}^0_{000}&=1+(\boostO-3)(\boostO-2)\frac{\langle p^2\rangle}{3}\approx 1+\The \boostO(\boostO-5)+6\The
\\
{}^{-3}\hat{\mathcal{D}}^0_{010}&=-\boostO(\boostO-5)\frac{\langle p^2\rangle}{3}\approx - \The \boostO(\boostO-5)
\end{align}
%---------------------------
are those that are relevant, giving the lab frame result
%---------------------------
\begin{align}
\label{app:col_term_int_K_lab}
\frac{\text{d}f_0}{\text{d}\tau}  
&\approx
\bar{\sigma}_1 \omega^2_1 \left[\Bigg\{{}^{-3}\hat{\mathcal{D}}^0_{000}+\frac{\alpha_{LR}}{3}{}^{-3}\hat{\mathcal{D}}^0_{010}- \sum_\ell {}^{-3}\hat{\mathcal{D}}^0_{0\ell 0}\Bigg\} f_0-\omega_1 \lambda_{LR}\,[1-2f_0] \, \hat{\mathcal{O}}_{\omega}f_0 +6 \omega_1 \lambda_{LR}\,f_0[1-f_0]\right]
\nonumber\\
&=\bar{\sigma}_1 \omega^2_1 \The\,\lambda_{LR}\left\{ \boostO(\boostO-5) f_0
-x [1-2f_0] \, \hat{\mathcal{O}}_{\omega}f_0 +6 x f_0[1-f_0]\right\},
\end{align}
%---------------------------
where $\sum_\ell {}^{-3}\hat{\mathcal{D}}^0_{0\ell 0}\approx {}^{-3}\hat{\mathcal{D}}^0_{000}+{}^{-3}\hat{\mathcal{D}}^0_{010}\approx 1+6\The$. Comparing to Eq.~\eqref{eq:FP} confirms the result and shows how the boost operator approach can be readily applied to the neutrino scattering problem.

%-------------------------------------------------
\subsection{Extension to other neutrino scattering processes}
\label{sec:extensions}
%-------------------------------------------------
The derivation above focused on $\nu_{\rm e}-e$ scattering. However, the same expressions apply to $\nu_{\mu/\tau}-e$ and the respective positron scattering processes when switching the relevant coefficient $\lambda_{LR}=1-\alpha_{LR}/3$ and scaling $\bar{\sigma}_1\propto (g_L^2+g_R^2-g_L g_R)$ \citep{Chluba2026nuKernels}. For the scattering of neutrinos with neutrons and protons, similar mappings appear \citep{Burrows2002,Suwa2019nuKomp}.

%-------------------------------------------------
\begin{table*}
\centering
\caption{Mapping of the leading Kompaneets coefficients for elastic
neutrino scattering off free electrons, positrons, protons and neutrons.
For all cases we write
$\sigma_s(E_\nu)=\sigma_{s,1}(E_\nu/\me)^2$ and define
$\lambda_s=1-a_s/3$, where $s=\{e^-, e^+, p, n \}$.
The ratios $R_\sigma$ and $R_K$ are defined relative to
$\nu_e-e^-$ scattering by
$R_\sigma=\sigma_{s,1}/\bar{\sigma}_1^{\nu_e e}$
and
$R_K=\sigma_{s,1}\lambda_s/
(\bar{\sigma}_1^{\nu_e e}\lambda_{\nu_e e})$, respectively.
For reference, $\bar{\sigma}_1^{\nu_e e}\approx \pot{1.11}{-20}\,\sigT$ and $\bar{\sigma}_1^{\nu_e e}\lambda_{\nu_e e}\approx \pot{0.96}{-20}\,\sigT$.
We also use $s^2_W=\sin^2\theta_W\approx 0.2312$ and $g_A\approx 1.2723$.
}
\label{tab:kompaneets_mapping}
\begin{tabular}{lcccccc}
\hline
Process
& Couplings & $a_s$ & $\lambda_s$ & $R_\sigma$ & $R_K$
\\
\hline
$\nu_e e^-$ &$\displaystyle
g_L=s_W^2+\frac12,\quad
g_R=s_W^2$
& $\alpha_{LR}=0.4031$
& $0.8656$
& $1$
& $1$
\\
$\nu_e e^+$
&
$\displaystyle
g_L=s_W^2,\quad
g_R=s_W^2+\frac12$
& $\alpha_{LR}=0.4031$
& $0.8656$
& $1$
& $1$
\\
$\nu_{\mu,\tau}e^-$
&
$\displaystyle
g_L=s_W^2-\frac12,\quad
g_R=s_W^2$
& $\alpha_{LR}=-0.3308$
& $1.1103$
& $0.4485$
& $0.575$
\\
$\nu_{\mu,\tau}e^+$
&
$\displaystyle
g_L=s_W^2,\quad
g_R=s_W^2-\frac12$
& $\alpha_{LR}=-0.3308$
& $1.1103$
& $0.4485$
& $0.575$
\\
$\nu p$
&
$\displaystyle
F_1^p=\frac{1-4s_W^2}{2},
\quad
F_A^p=\frac{g_A}{2}$
& $\delta_p\simeq-0.33$
& $1.11$
& $0.726$
& $0.930$
\\
$\nu n$
&
$\displaystyle
F_1^n=-\frac12,
\quad
F_A^n=-\frac{g_A}{2}$
& $\delta_n\simeq-0.11$
& $1.037$
& $0.874$
& $1.046$
\\
\hline
\end{tabular}
\end{table*}
%-------------------------------------------------

The mappings for the various scattering channels are summarized in Table~\ref{tab:kompaneets_mapping}. 
In all cases, the FP limit of the evolution equation can be written as
%-------------------------------------------------
\bealf{
\label{eq:Komp_gamma_nu_s}
\frac{\partial f}{\partial y_s}&=\frac{1}{x^2}\frac{\partial}{\partial x} x^6 \left[\frac{\partial f}{\partial x}+\frac{T_\nu}{T_s} f(1-f)\right]
}
%-------------------------------------------------
with the corresponding Kompaneets time variable ${\rm d}y_s
= N_s\,\sigma_{s,1}\lambda_s
\left(\frac{T_s}{m_s}\right)
\left(\frac{T_\nu}{\me}\right)^2{\rm d}t$, where $N_s$ denotes the number density of the scatter and $T_s/\hat{\mu}_s$ is the dimensionless temperature of the scatter in units of the scatterers rest mass.

For the neutrino-electron and -positron scattering processes, in all cases the overall effective cross section is
$\sigma_{s, 1}\lambda_s=\frac{4G_F^2\me^2}{\pi}
\left(g_L^2+g_R^2-\frac{4}{3}g_Lg_R
\right)$ with the respective coupling constants $g_L$ and $g_R$. Since $\sigma_{s, 1}\lambda_s$ is symmetric in the exchange $g_L\leftrightarrow g_R$ the neutrino-electron and neutrino-positron scattering processes have the same Kompaneets variable, $y_s$. Relative to $\nu_e-e^\pm$ scattering the scattering efficiency for $\nu_{\mu,\tau}-e^\pm$ scattering is thus about $40\%$ lower.

For neutrino-proton scattering we have $\sigma_{s, 1}\lambda_s=
\frac{2G_F^2 \me^2}{3\pi}
\left[(F_1^p)^2+5(F_A^p)^2\right]$ and similarly for the neutrino-neutron scattering process but with $F_1^p\rightarrow F_1^n$ and $F_A^p\rightarrow F_A^n$. Again comparing to $\nu_e-e^\pm$ scattering, the $\nu-n/p$ scattering efficiency is overall similar when assuming the same particle number density and dimensionless temperature. We note that due to the large mass of the nucleon, at the same temperature one has $T_s/m_s\ll \Te/\me$ and thus the validity of the FP limit is improved.

%-------------------------------------------------
\section{Approximate solutions to the photon and neutrino Kompaneets equations}
\label{sec:analytics}
%-------------------------------------------------
We now consider the Kompaneets equation of neutrinos and photons in limiting cases. For this it is first useful to write all equations in terms of $x=E/T_{\rm r}$ with reference temperature $T_{\rm r}=T_\nu$ for the neutrinos and $T_{\rm r}=T_\gamma$ for photons. To distinguish the two cases, we will use $f$ for the neutrino case and $n$ to denote the photon occupation number. This then gives the two evolution equations
%-------------------------------------------------
\bealf{
\label{eq:Komp_gamma_nu}
\frac{\partial n}{\partial y_\gamma}&=\frac{1}{x^2}\frac{\partial}{\partial x} x^4 \left[\frac{\partial n}{\partial x}+\frac{\Tg}{\Te} n(1+n)\right],
&
\frac{\partial f}{\partial y_\nu}&=\frac{1}{x^2}\frac{\partial}{\partial x} x^6 \left[\frac{\partial f}{\partial x}+\frac{T_\nu}{\Te} f(1-f)\right].
}
%-------------------------------------------------
For convenience we introduced the respective scattering $y$-parameters as a new time variable through
%-------------------------------------------------
\bealf{
\id y_\gamma &= \The \, \Ne \sigT \id t, 
&
\id y_\nu &= \The\,\theta_\nu^2 \, \Ne \bar{\sigma}_1\lambda_{LR} \id t. 
}
%-------------------------------------------------
This already indicates that the scattering efficiency for neutrinos has a stronger temperature dependence than for photons. In addition, since $\bar{\sigma}_1\ll \sigT$, a much higher electron density is required to have the same scattering rate. 
Conversely, to obtain the same value for the scattering $y$-parameter at fixed electron number density one therefore has to integrate over a much longer time-interval.

%-------------------------------------------------
\subsection{Properties of the Kompaneets operator and the equilibrium temperature}
%-------------------------------------------------
The Kompaneets operators in Eq.~\eqref{eq:Komp_gamma_nu} directly show that the scattering process does not change the number of photons or neutrinos. This can be easily seen when integrating the evolution equations over $x^2 \id x$ and using that for physical systems the respective distributions vanish at the boundaries. 

One can also easily show that the equilibrium distributions are $n_{\rm BE}=1/(\expf{x+\mu}-1)$ with constant chemical potential for the photons and $f_{\rm FD}=1/(\expf{x+\mu}+1)$ for the neutrinos. These solutions are approached naturally in the limit of many scatterings, showing that without number changing processes full thermal equilibrium (i.e., $\mu=0$) cannot be reached. 

In a similar way, we can treat the energy transfer with the electrons, by integrating the evolution equations in Eq.~\eqref{eq:Komp_gamma_nu} over $x^3 \id x$.
For the photons with energy density $\rho_\gamma\propto \int x^3  n \id x$, we can write
%-------------------------------------------------
\bealf{
\label{eq:Te_C}
\frac{\partial \rho_\gamma}{\partial y_\gamma}&=4 \rho_\gamma \left[1-\frac{\Te^{\rm C}}{\Te}\right],
\qquad\qquad
\Te^{\rm C}=\Tg \frac{\int x^4  n(1+n) \id x}{4\int x^3  n \id x},
}
%-------------------------------------------------
where $\Te^{\rm C}$ is the Compton equilibrium temperature that is approached in the given radiation field for many scatterings \citep{Zeldovich1970TCompton, Sazonov2001}. For $\Te=\Te^{\rm C}$, the net energy exchange between the (distorted) photon field and the electrons vanishes, while photons can in principle still be redistributed in energy. For Planckian photons, i.e, $n_{\rm eq}=1/(\expf{x}-1)$, one has $\partial_x n_{\rm eq} = -n_{\rm eq}(1+n_{\rm eq})$ and thus $\Te^{\rm C}=\Tg$, such that full equilibrium is reached when $\Te=\Tg$. Similarly, for a Bose-Einstein spectrum, $n_{\rm BE}=1/(\expf{x+\mu}-1)$ with constant chemical potential, $\mu$, one finds $\partial_x n_{\rm BE} = -n_{\rm BE}(1+n_{\rm BE})$ and $\Te^{\rm C}=\Tg$, as is well-known in connection with CMB spectral distortion physics \citep{Burigana1991,Hu1993}.

For neutrinos, the equilibrium temperature similarly follows by integrating the neutrino Kompaneets equation over $x^3 \id x$. With $\rho_\nu\propto \int x^3  f \id x$ for the neutrino energy density this yields 
%-------------------------------------------------
\bealf{
\label{eq:Te_C_nu}
\frac{\partial \rho_\nu}{\partial y_\nu}&=6 \rho_\nu \,\epsilon_\nu\,\left[1-\frac{\Te^{\rm C\nu}}{\Te}\right],
\qquad\qquad
\Te^{\rm C\nu}=T_\nu \frac{\int x^6  f(1-f) \id x}{6\int x^5  f \id x},
}
%-------------------------------------------------
where $\Te^{\rm C\nu}$ is the equilibrium temperature in the given neutrino field, which again determines the stationary solution. 
We note that the extra factor of $x^2$ comes from the energy-dependence of the neutrino-electron scattering cross section. This also causes the new factor 
%-------------------------------------------------
\bealf{
\label{eq:epsilon_nu}
\epsilon_\nu=
\frac{\int \sigma_\nu (x) \, x^3  f \id x}{\int \sigma_\nu (x=1)\, x^3  f \id x}
\equiv \frac{\int x^5  f \id x}{\int x^3  f \id x},
}
%-------------------------------------------------
which we shall refer to as the energy-density weighted effective cross section enhancement factor. For photons, this factor is $\epsilon_\gamma=1$ since the cross section is energy independent in the Kompaneets limit. For mono-energetic neutrinos, one has $\epsilon_\nu=x^2$, immediately making the physical meaning of $\epsilon_\nu$ clear. 
For neutrino distributions close to full equilibrium, one has $\epsilon_\nu \approx 20.81$, where we used the integrals $\mathcal{G}_k^\nu=\int x^k / (\expf{x}+1)\id x=(1-2^{-k})\, k! \,\zeta(k+1)$ in terms of the Riemann $\zeta$ function. Note that one also has 
$\mathcal{G}_k^\nu\equiv (1-2^{-k})\,\mathcal{G}_k^\gamma$ with $\mathcal{G}_k^\gamma=\int x^k / (\expf{x}-1)\id x=k! \,\zeta(k+1)$.

%-------------------------------------------------
\subsection{Doppler-dominated scattering}
\label{sec:Doppler_dom_sol}
%-------------------------------------------------
We now consider the case when the electron distribution is much hotter than the equilibrium temperature. That means that the recoil terms can be neglected, giving scattering problems of the form
%-------------------------------------------------
\bealf{
\label{eq:Komp_gamma_nu_Dop_limit}
\frac{\partial n}{\partial y_\gamma}&=\frac{1}{x^2}\frac{\partial}{\partial x} x^4 \frac{\partial n}{\partial x},
&
\frac{\partial f}{\partial y_\nu}&=\frac{1}{x^2}\frac{\partial}{\partial x} x^6 \frac{\partial f}{\partial x}.
}
%-------------------------------------------------
This is the so-called Doppler-dominated scattering regime, where energy is transfered to the photon or neutrino field. For the photon case, the solution can be written as \citep{Zeldovich1969}
%-------------------------------------------------
\bealf{
\label{eq:Greens_Dopp_gamma}
n(y_\gamma, x)&=\int \frac{n(0, x')}{\sqrt{4\pi y_\gamma}}\,\exp\left(-\frac{[\ln(x/x')+3y_\gamma]^2}{4y_\gamma}\right)\,\frac{\!\id x'}{x'},
}
%-------------------------------------------------
with initial condition $n(0, x)$. This solution can be readily found by transforming to $\xi=\ln x$ and $s=x^3 n$, which then gives $\partial_{y_\gamma} s = \partial_\xi^2 s - 3 \partial_\xi s$. Further using $z=\xi - 3 y_\gamma$ then yields $\partial_{y_\gamma} s = \partial_z^2 s$ and thus the diffusion solution given above once transformed back to $n(y_\gamma, x)$.

Assuming a number-normalized line, $n(0, x)=N_0 \delta(x-x_0)/x^2$, we can show that the position of the line center in terms of specific photon number ($N_x \propto x^2 n$) and intensity ($I_x \propto x^3 n$) moves as \citep{Chluba2008d}
%-------------------------------------------------
\bealf{
x^\gamma_N(y_\gamma)&=x_0\,\exp(y_\gamma),
&
x^\gamma_I(y_\gamma)&=x_0\,\exp(3y_\gamma),
}
%-------------------------------------------------
showing a net increase by scattering. The average energy per injected photon then evolves as %-------------------------------------------------
\bealf{
\langle x\rangle_\gamma=\frac{\int x^3 n \id x}{\int x^2 n \id x}=x_0 \,\exp(4y_\gamma),
}
%-------------------------------------------------
illustrating the link to the energy transfer per scattering as $\Delta x/x\approx 4\The$ (i.e., the first energy moment).

For the neutrino case, we can use $r=1/x$ and $u=f/r^3$. Inserting this into the neutrino Kompaneets equation, one finds 
%-------------------------------------------------
\bealf{
\label{eq:Komp_nu_Doppler}
\frac{\partial f}{\partial y_\nu}&=r^3 \frac{\partial u}{\partial y_\nu}\equiv
r^2 \frac{\partial}{\partial x} x^6 \frac{\partial r^3 u}{\partial x}=r^2\frac{1}{x^2} \frac{\partial}{\partial r} x^6 \frac{1}{x^2}\frac{\partial r^3 u}{\partial r}
=r^4 \frac{\partial}{\partial r}\frac{1}{r^4} \frac{\partial r^3 u}{\partial r}=r^4 \frac{\partial}{\partial r}\frac{1}{r^4} \left[3r^2 u+r^3\frac{\partial u}{\partial r}\right]
\nonumber\\
&=r^4 \frac{\partial}{\partial r}\left[\frac{3 u}{r^2}+\frac{1}{r}\frac{\partial u}{\partial r}\right]=r^4 \left[\frac{3 \partial_r u}{r^2}-\frac{6u}{r^3}+\frac{\partial^2_r u}{r} -\frac{\partial_r u}{r^2}\right]=r^3 \partial^2_r u+2 r^2 \partial_r u - 6r u,
}
%-------------------------------------------------
implying $\partial_{y_\nu} u = \partial_r^2 u+2\partial_r u/r-6u/r^2$. This equation can be solved by separating variables with the Ansatz $u=T(y_\nu) \phi(r)$, which yields $\partial_{y_\nu}T=-k^2 T$ and $\partial_r^2\phi+(2/r)\partial_r\phi+(k^2-6/r^2)\phi=0$ with the constant $k$. The equation for $T$ has the solution $T(y_\nu)=\expf{-k^2y_\nu}$. The expression for $\phi$ is solved by spherical Bessel functions with $\ell = 2$. This means that the regular solution is $\phi(r) = A(k) j_2(k r)$ and hence
$u(y_\nu, r)=\int \id k\, A(k)\, \expf{-k^2 y_\nu} j_2(k r)$.

To determine $A(k)$, we can use the initial condition $u(0, r)=\int \id k\, A(k) j_2(k r)$. Multiplying by $r^2 j_2(k' r)$ and integrating over $\id r$ we then have
%-------------------------------------------------
\bealf{
\int r^2 j_2(k' r) \, u(0, r) \id r = \int r^2 j_2(k' r)\int \id k\, A(k) j_2(k r) = \frac{\pi \,A(k')}{2k'^2},
}
%-------------------------------------------------
where in the last step we used the orthogonality condition for the spherical Bessel function. This then gives the overall solution
%-------------------------------------------------
\bealf{
u(y_\nu, r)&=\int u(0, r') \,K(y_\nu, r', r) \,r'^2\id r', 
\nonumber\\
K(y, r', r)&=\frac{2}{\pi}\int \id k\, k^2 \, \expf{-k^2 y} j_2(k r) j_2(k r')
\nonumber\\
&=\frac{1}{\sqrt{4\pi y}\,rr'}\left[\left(1-\frac{6y}{rr'}+\frac{12y^2}{r^2 r'^2}\right)\expf{-\frac{(r-r')^2}{4y}}
-\left(1+\frac{6y}{rr'}+\frac{12y^2}{r^2 r'^2}\right)\expf{-\frac{(r+r')^2}{4y}}\right],
}
%-------------------------------------------------
where the kernel integral can be readily taken with {\tt Mathematica}. Transforming back to $x=1/r$ and $f= r^3 u=u/x^3$ with $\id r'=-r'^2\id x'$, we then have the final solution
%-------------------------------------------------
\bealf{
\label{eq:Greens_Dopp_nu}
f(y_\nu, x)&=\int \frac{f(0, x')}{\sqrt{4\pi y_\nu}x^2}\left\{\left[1-6 x x' y_\nu +12 (x x'y_\nu)^2\right]\expf{-\frac{(x'-x)^2}{4y_\nu x^2 x'^2}}
-\left[1+6 x x' y_\nu +12 (x x'y_\nu)^2\right]\expf{-\frac{(x'+x)^2}{4y_\nu x^2 x'^2}}\right\}\!\id x'
\nonumber\\
&\stackrel{\stackrel{y_\nu \ll 1}{\downarrow}}{\approx}\int \frac{f(0, x')}{\sqrt{4\pi y_\nu}x^2}\left[1-6 x x' y_\nu +12 (x x'y_\nu)^2\right]\expf{-\frac{(x'-x)^2}{4y_\nu x^2 x'^2}}
\!\id x'
}
%-------------------------------------------------
in terms of elementary functions. In the last line we assumed that $y_\nu\ll 1$ and that one does not go too far into the high energy wing of the scattered distribution\footnote{We find the approximation to work very well for $x_0^2 y \lesssim 0.01$ and $x\lesssim 20 x_0$.}. This highlights the asymmetry of the profile due to the exponential factor and overall weighting function. For a number-normalized initial line, $f(0, x)=N_0 \delta(x-x_0)/x^2$, we can again determine the average energy per injected neutrino. Defining $\epsilon \equiv 4y_\nu x_0^2$, we find %-------------------------------------------------
\bealf{
\langle x\rangle_\nu=\frac{\int x^3 f \id x}{\int x^2 f \id x}
%=x_0 \,
%\frac{3\sqrt{\pi}}{4\epsilon}\,
%\frac{
%\left(4+4\epsilon+3\epsilon^2\right)
%D\!\left(\frac{1}{\sqrt{\epsilon}}\right)
%-\sqrt{\epsilon}\left(2+3\epsilon\right)
%}{
%3\sqrt{\pi}\,\epsilon^{3/2}
%\operatorname{erf}\!\left(\frac{1}%{\sqrt{\epsilon}}\right)
%-2\left(2+3\epsilon\right)\expf{-1/\epsilon}
%}
%
%\frac{3\sqrt{\pi}}{16}
%\,
%\frac{
%(4+4\epsilon+3\epsilon^2)\frac{1}{\sqrt{\epsilon}}
%D\!\left(\frac{1}{\sqrt{\epsilon}}\right)
%-\left(2+3\epsilon\right)
%}{\epsilon^2\,\Gamma\left(\frac{5}{2},\frac{1}{\epsilon}\right)}
%
\stackrel{\stackrel{\epsilon\ll1}{\downarrow}}{\approx} x_0\left(1+\frac{3}{2}\epsilon+\frac{9}{2}\epsilon^2\right),
}
%-------------------------------------------------
%in terms of the Dawson function, $D(z)$, and incomplete $\Gamma$ function, $\Gamma(a,z)$.
%
which again highlights the link to the energy transfer per scattering, $\Delta x/x\approx \frac{3}{2}\frac{\The \epsilon}{y_\nu x_0^2}=
6\The$ for $y_\nu\ll 1$. Similarly, we find $x^\nu_N(y_\nu)\approx x_0 (1-3\epsilon^2/4)$ and $x^\nu_I(y_\nu)\approx x_0 (1+\epsilon/2-\epsilon^2/4)$. Since recoil was not included, one has to take proper care when considering the contributions at $x\rightarrow \infty$, as neutrinos reach the boundary with a finite probability due to this approximation. We discuss the effect of this in Section~\ref{sec:line_inj}, Figs.~\ref{fig:line_evo_Kompaneets_I} and \ref{fig:line_evo_Kompaneets_II}.

%-------------------------------------------------
\subsection{Recoil-dominated scattering regime}
%-------------------------------------------------
Next we consider the solution in the recoil-dominated regime, neglecting any Pauli-blocking or stimulated scattering effects. Setting $\theta_\nu=\The$, the evolution equations then read
%-------------------------------------------------
\bsub
\bealf{
\frac{\partial n}{\partial y_\gamma}&\approx \frac{1}{x^2}\frac{\partial}{\partial x} x^4 n, 
&
\frac{\partial f}{\partial y_\nu}&\approx \frac{1}{x^2}\frac{\partial}{\partial x} x^6 f,
}
\esub
%-------------------------------------------------
which applies in regimes when the electrons have much lower energies than the photon or neutrino. 
Using characteristics, the solutions read
%-------------------------------------------------
\bsub
\label{eq:sol_recoil_only}
\bealf{
n(y_\gamma, x)&=\frac{1}{(1-y_\gamma x)^4}\,n\left(0, \frac{x}{1-y_\gamma x}\right), 
&
f(y_\nu, x)&=\frac{1}{(1-3y_\nu x^3)^2}\,f\left(0, \frac{x}{(1-3y_\nu x^3)^{1/3}}\right), 
}
\esub
%-------------------------------------------------
where the solution for the photons was given by \citep{Arons1971, Illarionov1972}. In both cases, an initial narrow line therefore simply moves down in energy without capturing the additional broadening effect (see section~\ref{sec:line_inj}). We note that for $\Te={\rm const}$ one has $y_\gamma x = \tau_\gamma \omega$ and similarly $y_\nu x^3 = \tau_\nu \omega^3$ with $\id \tau_\gamma = \id y_\gamma/\The$ and $\id \tau_\nu = \id y_\nu/\The^3$ determining the related optical depth variables, i.e., independent of temperature.

\paragraph{Non-linear recoil with stimulated scattering}
%-----------------
For photons, one can also consider the case of non-linear photon evolution due to stimulated recoil \citep{Levich1969, Zeldovich1972shock}. The relevant evolution equation then is \citep{Zeldovich1972shock} $\frac{\partial n}{\partial y_\gamma}\approx \frac{1}{x^2}\frac{\partial}{\partial x} x^4 n^2$ and can be solved implicitly \citep[see also][]{Chluba2008d}. For neutrinos, $f<1$ and the equivalent situation is when $f\rightarrow 1$ such that Pauli-blocking is extremely efficient and the redistribution by scattering is entirely stopped. This situation only appears for very dense states not relevant to astrophysical/cosmological plasmas where in equilibrium one usually has $f\rightarrow 1/2$ at $x\ll 1$. We therefore do not consider this case any further. 

We remark though that the term $1-f$ in the neutrino Kompaneets equation is crucial for estabilishing full equilibrium under repeated scattering. The stationary solution to Eq.~\eqref{eq:Komp_gamma_nu} is determined by $\partial_x f = -\frac{T_\nu}{\Te} f(1-f)$, which yields $f_{\rm FD}=1/(\expf{x-m}+1)$, where $m$ is the required integration constant to fix the number of neutrinos in the system (i.e., the dimensionless chemical potential) for $T_\nu=\Te$. Without the factor $1-f$ one would find $f_{\rm W}\approx \expf{-x+m}$ which omits quantum effects.

%-------------------------------------------------
\subsection{Doppler-dominated with background-induced scattering}
%-------------------------------------------------
Another case of interest, is when considering the evolution of a low energy ($x\lesssim 1$) distortion feature on a thermal background with $T_{\rm r}=\Te$. Using $n=n_{\rm eq}+\Delta n$ and $f=f_{\rm eq}+\Delta f$ and linearizing in the perturbation, one finds
%-------------------------------------------------
\bealf{
\frac{\partial \Delta n}{\partial y_\gamma}&\approx \frac{1}{x^2}\frac{\partial}{\partial x} x^4 \left[\frac{\partial \Delta n}{\partial x}+\Delta n(1+2 n_{\rm eq})\right],
&
\frac{\partial \Delta f}{\partial y_\nu}&\approx \frac{1}{x^2}\frac{\partial}{\partial x} x^6 \left[\frac{\partial \Delta f}{\partial x}+\Delta f(1-2 f_{\rm eq})\right],
}
%-------------------------------------------------
where now $x=E/\Te$ such that $n_{\rm eq}=1/(\expf{x}-1)$ and $f_{\rm eq}=1/(\expf{x}+1)$. Further using
%------------------------
\begin{align}
A=
\begin{cases}
1+ 2 n_{\rm eq}\approx  \frac{2}{x} &{\rm bosons}
\\
1- 2 f_{\rm eq}\approx 0 &{\rm fermions},
\end{cases}
\end{align}
%------------------------
for $x\ll 1$, we have 
%-------------------------------------------------
\bealf{
\frac{\partial \Delta n}{\partial y_\gamma}&\approx \frac{1}{x^2}\frac{\partial}{\partial x} x^4 \left[\frac{\partial \Delta n}{\partial x}+\frac{2}{x} \Delta n \right],
&
\frac{\partial \Delta f}{\partial y_\nu}&\approx \frac{1}{x^2}\frac{\partial}{\partial x} x^6 \frac{\partial \Delta f}{\partial x}.
}
%-------------------------------------------------
This implies that there is no leading order background-induced effect for the scattering of a low-energy neutrino distortions in an ambient thermal neutrino background in equilibrium with the electrons, and the solution in Eq.~\eqref{eq:Greens_Dopp_nu} applies. For photons on the other hand one finds
\citep{Chluba2008d}
%-------------------------------------------------
\bealf{
\label{eq:Greens_Dopp_gamma_stim}
n(y_\gamma, x)&=\int \frac{x'^3}{x^3}\,\frac{n(0, x')}{\sqrt{4\pi y_\gamma}}\,\exp\left(-\frac{[\ln(x/x')-y_\gamma]^2}{4y_\gamma}\right)\,\frac{\!\id x'}{x'}.
}
%-------------------------------------------------
In this case one has
%-------------------------------------------------
\bealf{
x^\gamma_N(y_\gamma)&=x_0\,\exp(-y_\gamma),
&
x^\gamma_I(y_\gamma)&=x_0\,\exp(y_\gamma),
&
\langle x\rangle&=x_0\,\exp(2y_\gamma)
}
%-------------------------------------------------
which highlights that stimulated scattering effects slow the up-scattering of photons by a factor $\simeq \exp(-2y_\gamma)$ with respect to the vacuum case \citep{Chluba2008d}, an effect that is absent for neutrinos. 

%-------------------------------------------------
\section{The neutrino relativistic SZ operator}
\label{sec:nuSZ_effect}
%-------------------------------------------------
For completeness we also give the scattering operator that allows us to compute the neutrino equivalent of the relativistic SZ effect \citep{Wright1979, Rephaeli1995, Sazonov1998, Itoh98, Challinor1998, Chluba2012SZpack} for an incoming background of neutrinos by relativistic thermal non-degenerate electrons. This operator can be obtained using the boost operator result that we already given in Sect.~\ref{sec:FP_nu} and provides a useful limiting description for background neutrinos traversing hot media.

%-------------------------------------------------
\subsection{Derivation of the relativistic SZ operator}
%-------------------------------------------------
The result in essentially given by the (thermally-averaged) equation 
%---------------------------
\begin{align}
\label{eq:SZ}
\frac{\text{d}f_0}{\text{d}\tau}  
&\approx
\bar{\sigma}_1 \omega^2_1 \Bigg\{{}^{-3}\hat{\mathcal{D}}^0_{000}+\frac{\alpha_{LR}}{3}{}^{-3}\hat{\mathcal{D}}^0_{010}- \sum_\ell {}^{-3}\hat{\mathcal{D}}^0_{0\ell 0}\Bigg\} \, f_0,
\end{align}
%---------------------------
which follows from Eq.~\eqref{app:col_term_int_K_lab} when neglecting recoil terms as would be appropriate when the electrons scattering the {\it isotropic} cosmic neutrino background have a much higher average energy. The key element is that the involved Doppler operators are exact to all orders in the electron momentum, $p$, and the related thermal averages can be carried out analytically for thermal electrons \citep[see][for related discussions with CMB photons]{Chluba2026SZ, Rosenberg2025prSZ, Hoey2026kSZ}.

To obtain the final result, we use the identity 
%---------------------------
\begin{align}
\label{eq:ID_I}
\sum_\ell {}^{-3}\hat{\mathcal{D}}^0_{0\ell 0}(p)&=\gamma^2(1+\beta^2)=1+2p^2,
\end{align}
%---------------------------
which follows from the properties of the boost operator and related Doppler operators \citep{ChlubaBO25, Hoey2026Kompaneets}. Using the method of \citep{Hoey2026Kompaneets, Hoey2026kSZ} it is also easy to show that
%---------------------------
\bsub
\label{eq:ID_II}
\begin{align}
^{-3}\hat{\mathcal{D}}^0_{000}(p)&=1+2p^2+
\sum_{n=1}^{\infty}
\frac{2^{2n+1}p^{2n}}{(2n+2)!}
\left[1+\frac{2(n+3)}{2n+3}\,p^2\right]\,\hat{P}_n,
\\
{}^{-3}\hat{\mathcal D}^{0}_{010}(p)
&=-\sum_{n=1}^{\infty}
\frac{3n}{(n+2)}\frac{2^{2n+1}p^{2n}}{(2n+2)!}
\left[1+\frac{2(n+3)}{2n+3}\,p^2\right]\,\hat{P}_n
\end{align}
\esub
%---------------------------
where we introduced the differential operators
%---------------------------
\label{eq:P_o}
\begin{align}
\hat{P}_n&=\prod_{m=0}^{n-1}\left[\boostO(\boostO-5)-m(m+5)\right]
\end{align}
%---------------------------
which are the main generators of the scattering-induced distortion. As examples, we have $\hat{P}_0=1$, $\hat{P}_1=\boostO(\boostO-5)$ and $\hat{P}_2=\boostO(\boostO-5)[\boostO(\boostO-5)-6]$. This then implies
%---------------------------
\bealf{
\label{eq:ID_III}
&{}^{-3}\hat{\mathcal D}^{0}_{000}
+\frac{\alpha_{LR}}{3}{}^{-3}\hat{\mathcal{D}}^0_{010}
-(1+2p^2)
=\sum_{n=1}^{\infty}
\left(1-\,\frac{\alpha_{LR}}{3}\frac{3n}{(n+2)}\right)\frac{2^{2n+1}p^{2n}}{(2n+2)!}
\left[1+\frac{2(n+3)}{2n+3}\,p^2\right]\,\hat{P}_n,
}
%---------------------------
which generates the scattering signal per electron momentum. 
To carry out the thermal average over non-degenerate relativistic electrons, one can use the momentum moments \citep[see][]{CSpack2019} 
%--------------------------------------------------------------
\bealf{
\label{app:MMrMB_k}
\left<p^{2n}\right>&=\frac{2\,(2\The)^{n} K_{n+2}(1/\The)}{\sqrt{\pi}K_2(1/\The)}\,\Gamma\left(n+\frac{3}{2}\right)
=
\frac{(2\The)^{n} K_{n+2}(1/\The)}{K_2(1/\The)}\,\frac{[2(n+1)]!}{2^{2n+1}(n+1)!},
}
%--------------------------------------------------------------
where $K_n(z)$ denotes the modified Bessel functions of second kind. 
This then allows another simplification by realizing that the thermal average of 
%--------------------------------------------------------------
\bealf{
\label{eq:A_n}
\langle A_n(p) \rangle &=\left<\frac{2^{2n+1}p^{2n}}{(2n+2)!}
\left[1+\frac{2(n+3)}{2n+3}\,p^2\right]\right>
\nonumber\\
&=
\frac{2^{2n+1}}{(2n+2)!}\left[\frac{(2\The)^{n} K_{n+2}(1/\The)}{K_2(1/\The)}\,\frac{[2(n+1)]!}{2^{2n+1}(n+1)!}+\frac{2(n+3)}{2n+3}\frac{(2\The)^{n+1} K_{n+3}(1/\The)}{K_2(1/\The)}\,\frac{[2(n+2)]!}{2^{2n+3}(n+2)!}\right]
\nonumber\\
&=
\frac{(2\The)^{n}}{(n+1)!}\frac{
\left[K_{n+2}(1/\The)
+ 2(n+3) \The K_{n+3}(1/\The)\right]}{K_2(1/\The)}
= \frac{(2\The)^{n}}{(n+1)!}
\frac{K_{n+4}(1/\The)}{K_2(1/\The)},
}
%--------------------------------------------------------------
due to identities for the modified Bessel functions.
This then gives the relativistic neutrino SZ-like operator
%---------------------------
\bealf{
\label{eq:S_SZ_nu}
\hat{\mathcal{S}}_{\rm \nu SZ}&=
\left<{}^{-3}\hat{\mathcal D}^{0}_{000}
+\frac{\alpha_{LR}}{3}{}^{-3}\hat{\mathcal{D}}^0_{010}
-(1+2p^2)\right>
=\sum_{n=1}^{\infty}
\left(1-\,\frac{\alpha_{LR}}{3}\frac{3n}{(n+2)}\right)\,\frac{(2\The)^{n}}{(n+1)!}
\frac{K_{n+4}(1/\The)}{K_2(1/\The)}\,\hat{P}_n.
}
%---------------------------
Expanding in terms of $\The$, one can then obtain an asymptotic expansion for the distortion caused by the scattering of the cosmic neutrino background by hot electrons. Defining $\bar{c}_n=\left(1-\,\frac{\alpha_{LR}}{3}\frac{3n}{(n+2)}\right)/\lambda_{LR}$, $\mathcal{F}_n=x^2 \hat{P}_n\,f_{\rm eq}$ with $f_{\rm eq}(x)=1/(\expf{x}+1)$, we can then write the neutrino distortion as
%--------------------------------------------------------------
\bsub
\label{eq:Dn_nu}
\bealf{
\Delta f_{\rm \nu SZ}&\approx y_\nu \left[\mathcal{Y}_{0}(x)+\The \mathcal{Y}_{1}(x)+\The^2 \mathcal{Y}_{2}(x)+\The^3 \mathcal{Y}_{3}(x)+\The^4 \mathcal{Y}_{4}(x)+\ldots\right]
\\
\mathcal{Y}_{0}(x)&=\mathcal F_1(x),
\\
\mathcal{Y}_{1}(x)&=\frac{21}{2}\mathcal F_1(x)
    + \frac{2}{3}\bar c_2\mathcal F_2(x),
\\
\mathcal{Y}_{2}(x)&=\frac{399}{8}\mathcal F_1(x)
    + \frac{32}{3}\bar c_2\mathcal F_2(x)
    + \frac{1}{3}\bar c_3\mathcal F_3(x),
\\
\mathcal{Y}_{3}(x)&=\frac{945}{8}\mathcal F_1(x)
    + 80\bar c_2\mathcal F_2(x)
    + \frac{15}{2}\bar c_3\mathcal F_3(x)
    + \frac{2}{15}\bar c_4\mathcal F_4(x),
\\
\mathcal{Y}_{4}(x)&=\frac{11655}{128}\mathcal F_1(x)
    + 340\bar c_2\mathcal F_2(x)
    + \frac{645}{8}\bar c_3\mathcal F_3(x)
    + 4\bar c_4\mathcal F_4(x)
    + \frac{2}{45}\bar c_5\mathcal F_5(x)
}
\esub
%--------------------------------------------------------------
with $x=\omega/\theta_\nu$ and where we assumed that $y_\nu = \int \Ne \,\The \theta_\nu^2 \,\bar{\sigma}_1 \lambda_{LR} \id t \ll 1$. It is easy to obtain the expression up to higher order in the electron temperature. However, like for the classical SZ effect \citep[see][for related discussion]{Itoh98, Chluba2012SZpack}, this expansion only converges slowly in the high energy part of the neutrino spectrum once the temperature exceeds $\The \simeq 0.03$ (see discussion around Fig.~\ref{fig:nu_SZ_signal}). If the scattering is dominated by nuclear reactions \citep[e.g., see][]{Suwa2019nuKomp}, then the involved particle momenta are significantly lower such that the related series is expected to converge more quickly.  

%-------------------------------------------------
\subsection{Spectral basis and illustration of the $\nu$SZ signal}
\label{sec:nuSZ_basis}
%-------------------------------------------------
The distortion caused by the scattering of background neutrinos by hot electrons is determined by the linear combination of the functions $\mathcal{F}_n(x)$. We can again learn from the photon-electron scattering problem, by recognizing that the function, $\mathcal{F}_n(x)$,  can be built by repeatedly applying the boost generator, $\boostO=-x\partial_x$ to the background distribution \citep{chluba_spectro-spatial_2023-I}. For CMB photons we have
%---------------------------------------
\begin{align}
\Gspec(x)
&=\boostO \nbb=\frac{x\expf{x}}{(\expf{x}-1)^2},
&
\Yspec(x)
&=\DiffO\nbb
=G\left[x\frac{\expf{x}+1}{\expf{x}-1}-4\right],
&
\Ynspec{k}(x)&=\frac{1}{4^k}\boostO^k \Yspec(x)
\end{align}
%---------------------------------------
using $\nbb=1/(\expf{x}-1)$ and where $\DiffO=\boostO(\boostO-3)=x^{-2}\partial_x x^4 \partial_x$ is the photon diffusion operator. This choice allows us to build all functions that are based on repeated boosts of $G(x)$ and $Y(x)$ \citep{chluba_spectro-spatial_2023-I}.

%-----------------------------------------------------
\begin{figure}
    \centering
    \includegraphics[width=0.76\columnwidth]{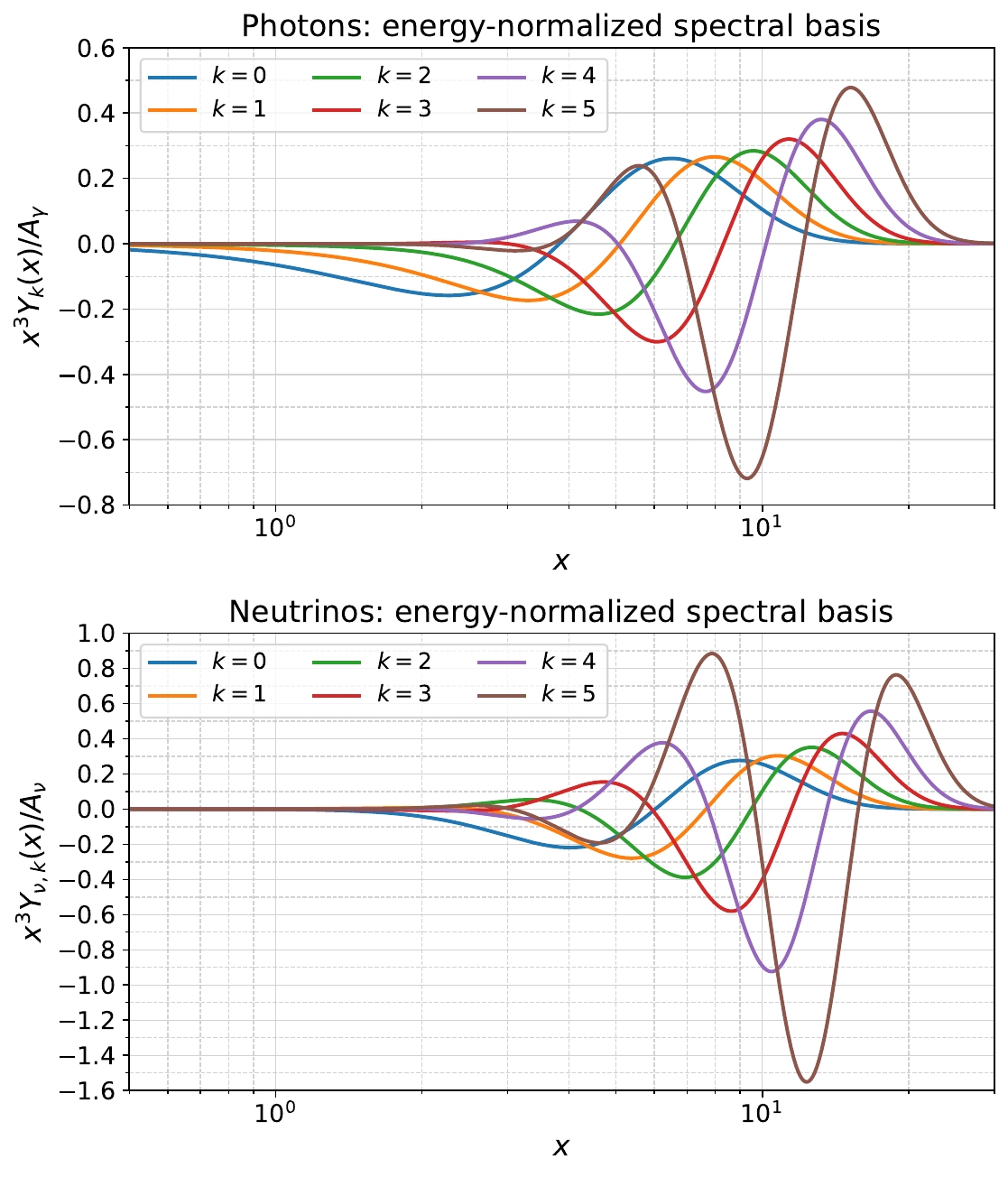}
    \vspace{-0mm}
    \caption{First few basis functions $\Ynspec{k}$ and $\Ynspec{k}^\nu$ normalized by $A_\gamma=4\pi^4/15$ and $A_\nu=31\pi^6/42$ to yield $\int x^3 \Ynspec{k}/A_\gamma \id x=\int x^3 \Ynspec{k}^\nu/A_\nu \id x=1$. The photon basis at a given $k$ is centered around lower energies than the neutrino basis. With increasing $k$ the basis functions shift focus towards higher frequencies.}
    \label{fig:nu_SZ_basis}
\end{figure}
%-----------------------------------------------------
In analogy to the photon case, we now introduce
%---------------------------------------
\begin{align}
\Gspec^\nu(x)
&=\boostO f_{\rm eq}=\frac{x\expf{x}}{(\expf{x}+1)^2},
&
\Yspec^\nu(x)
&=\DiffO^\nu f_{\rm eq}
=x^2 \Gspec^\nu\left[x\,\frac{\expf{x}-1}{\expf{x}+1}-6\right],
&
\Ynspec{k}^\nu(x)&=\frac{1}{4^k}\boostO^k \Yspec^\nu(x) 
\end{align}
%---------------------------------------
using $f_{\rm eq}=1/(\expf{x}+1)$ and where $\DiffO^\nu=x^2 \boostO(\boostO-5)=x^{-2}\partial_x x^6 \partial_x$ is the neutrino diffusion operator. How did we choose these? First of all, $\Yspec^\nu(x)=\DiffO^\nu\nbb$ and the $\Ynspec{k}^\nu(x)$ only carry energy, but no photon number, i.e., $\int x^2 \Yspec^\nu(x)\id x=\int x^2 \Ynspec{k}^\nu(x)\id x=0$ and\footnote{As part of the intermediate steps we used $\int x^\alpha \boostO^k f(x)\id x=(1+\alpha)^k \int x^\alpha f(x)\id x$  and $\boostO(\boostO-5) =(\boostO-5) \boostO$ \citep{Chluba2026}.}
%---------------------------------------
\bealf{
\int x^3 \Yspec^\nu(x)\id x&=\int x^5 \boostO(\boostO-5) f_{\rm eq} \id x= \int x^5 (\boostO-5) \boostO f_{\rm eq} \id x = \int x^5 \boostO f_{\rm eq} \id x
\nonumber\\
&=\int x^5 \Gspec^\nu(x) \id x =6\int x^5 f_{\rm eq} \id x = \frac{31 \pi^6}{42}\approx 709.60.
}
%---------------------------------------
The extra factor of $x^2$ in the definition of $\Yspec^\nu(x)$ stems from the cross section scaling and is crucial for ensuring that $\int x^2 \Yspec^\nu(x)\id x=0$. The functions $\Ynspec{k}^\nu(x)$ were furthermore scaled by $1/4^k$ since this keeps their energy densities constant $\int x^3 \Ynspec{k}^\nu(x)\id x\equiv \int x^3 \Yspec^\nu(x)\id x$. 

In Fig.~\ref{fig:nu_SZ_basis}, we illustrate the first few basis functions $\Ynspec{k}$ and $\Ynspec{k}^\nu$. To link the $\Ynspec{k}^\nu$ back to $\mathcal{F}_n$, we recognize that 
%---------------------------------------
\bealf{
\mathcal{F}_n&=x^2 \left\{\prod_{m=0}^{n-1}\left[\boostO(\boostO-5)-m(m+5)\right] \right\} f_{\rm eq}=x^2 \left\{\prod_{m=1}^{n-1}\left[\boostO(\boostO-5)-m(m+5)\right]\right\}x^{-2} \Yspec^\nu(x)
\nonumber\\
&=\left\{\prod_{m=1}^{n-1}\left[\boostO(\boostO-1)-(m+2)(m+3)\right]\right\} \Yspec^\nu(x)
=\left\{\prod_{m=3}^{n+1}\left[\boostO(\boostO-1)-m(m+1)\right]\right\} \Yspec^\nu(x),
}
%---------------------------------------
where we used $x^2 \boostO x^{-2}=\boostO+2$. This means %---------------------------------------
\bsub
\bealf{
\mathcal{F}_1&=\Ynspec{0}^\nu(x)=\Yspec^\nu(x)=\frac{x^3 \expf{x}}{(\expf{x}+1)^2} \left[x\,\frac{\expf{x}-1}{\expf{x}+1}-6\right], 
\\
\mathcal{F}_2&=-12\Ynspec{0}^\nu(x)-4\Ynspec{1}^\nu(x)+16 \Ynspec{2}^\nu(x)
\\
\mathcal{F}_3 &=
240\,\Ynspec{0}^\nu
+128\,\Ynspec{1}^\nu
-496\,\Ynspec{2}^\nu
-128\,\Ynspec{3}^\nu
+256\,\Ynspec{4}^\nu ,
\\
\mathcal{F}_4 &=
-7200\,\Ynspec{0}^\nu
-4800\,\Ynspec{1}^\nu
+18208\,\Ynspec{2}^\nu
+7872\,\Ynspec{3}^\nu
-15104\,\Ynspec{4}^\nu
-3072\,\Ynspec{5}^\nu
+4096\,\Ynspec{6}^\nu
}
\esub
%---------------------------------------
and so on, showing that the $\Ynspec{k}^\nu(x)$ provide an alterative basis.

%-----------------------------------------------------
\begin{figure}
    \centering
    \includegraphics[width=0.76\columnwidth]{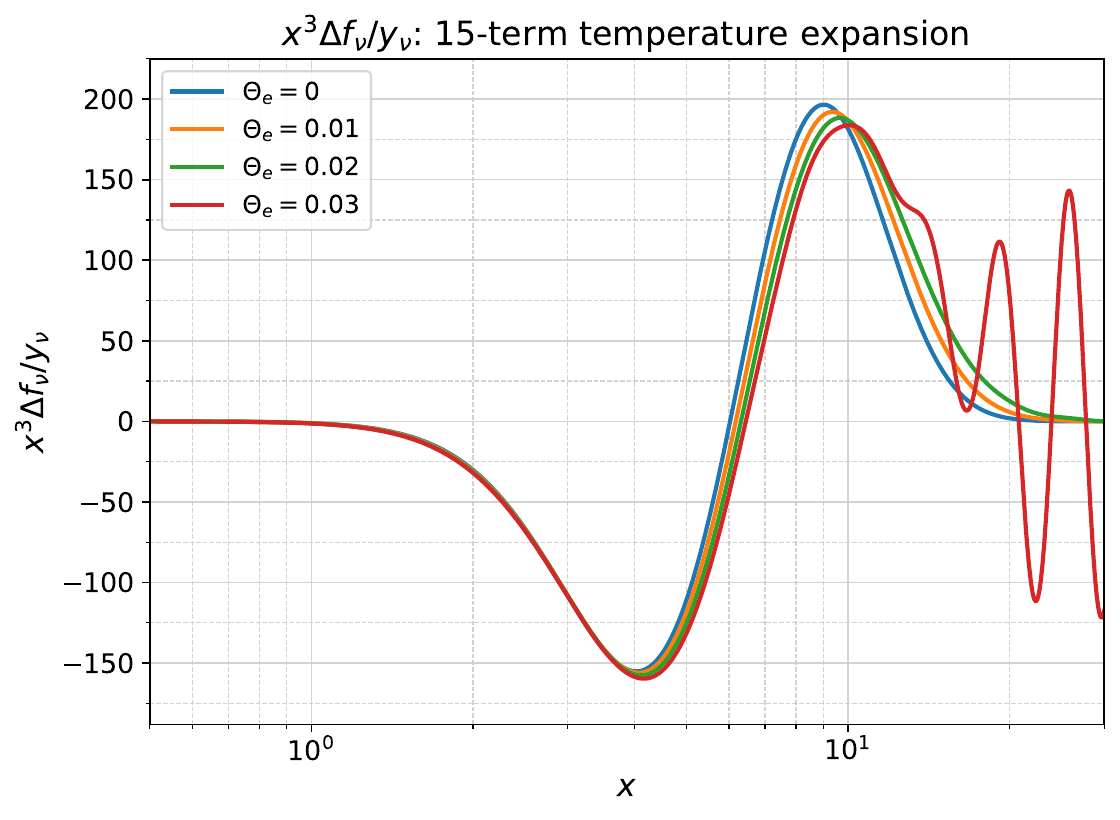}
    \vspace{-0mm}
    \caption{Scattering induced thermal $\nu$SZ signal for varying temperature of the scattering electrons. Increasing the temperature shifts the distortion shape upwards in energy. For $\The=0.03$ the 15-term expansion no longer converges at $x>10$, showing the asymptotic nature of the expansion like for the photon case.}
    \label{fig:nu_SZ_signal}
\end{figure}
%-----------------------------------------------------
In Fig.~\ref{fig:nu_SZ_signal} we illustrate the relativistic $\nu$SZ signal shape. Like in the photon case, the general trend is to shift the non-relativistic distortion signal, $\simeq  y_\nu\,x^3 \mathcal{F}_(x)$, upwards in energy with increasing temperature. However, unlike the CMB SZ effect, one can see that the low-frequency distortion stays roughly independent of temperature because the scattering cross section scales as $\sigma \propto E_\nu^2$ instead of being energy independent as for photons. In addition, we can see that the 15 term expansion does not converge at $x>10$ for $\The=0.03$, as expected from the asymptotic nature of the expansion. An exact scattering kernel evaluation avoids this limitation, as we will see below. For $\nu$SZ signals due to the scattering with nucleons, the same expressions as for $\nu_e-e$ scattering can be applied but with the replacements $\alpha_{LR}\rightarrow a_s$, $\lambda_{LR}\rightarrow 1-a_s/3$, $y_
\nu\rightarrow y_s$ and $\The \rightarrow T_s/m_s$ (see Sect.~\ref{sec:extensions}).
Since $\me \ll m_s$, we expect the convergence of the expressions to be improved for the scattering by nucleons.

%-----------------------------------------------------
\begin{figure}
    \centering
    \includegraphics[width=0.8\columnwidth]{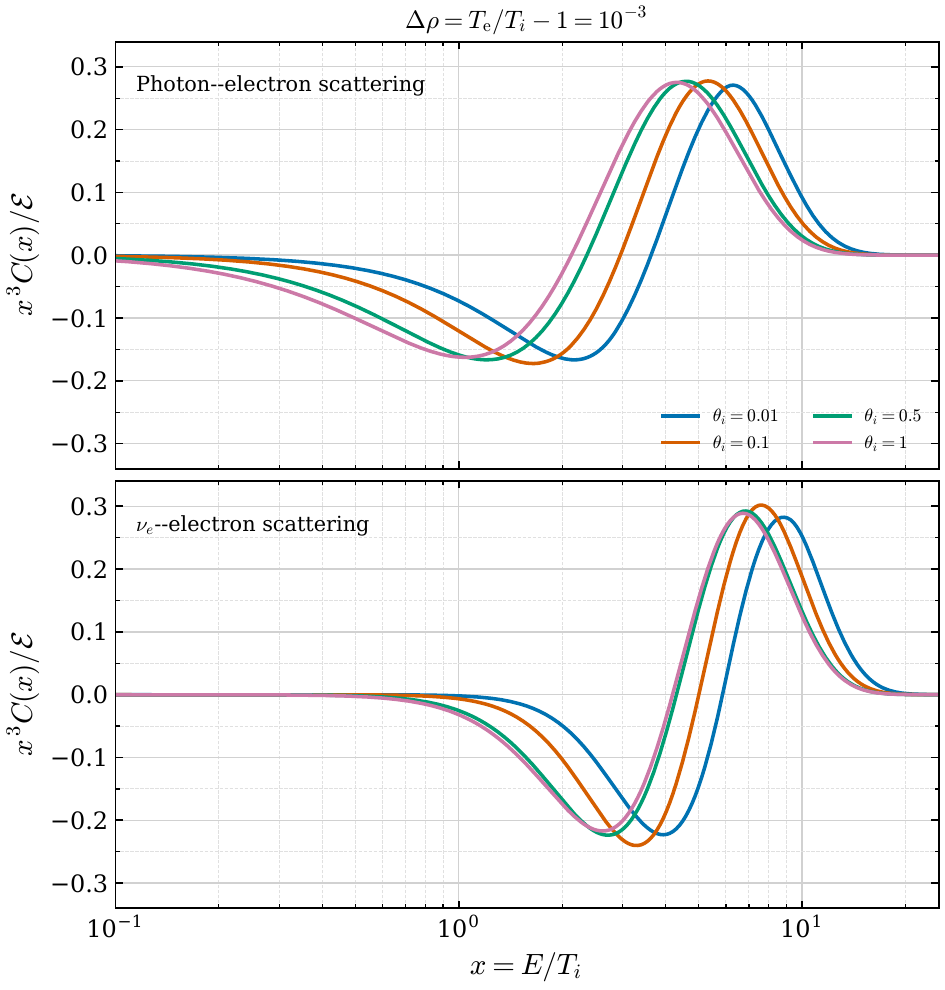}
    \vspace{-0mm}
    \caption{Collision term for photons (top) and neutrinos (bottom) scattering with electrons at varying temperatures $\theta_i$ (with $i=\{\gamma, \nu\}$) and $\Delta \rho=10^{-3}$. We scaled all cases by the energy exchange integral, $\mathcal{E}=\int x^3 \mathcal{C} \id x$, to normalize them. For increasing temperature, the importance of recoil increases as indicated by the shift of the null towards lower energies.}
    \label{fig:Coll_Drho_0.001}
\end{figure}
%-----------------------------------------------------
We also note that the non-relativistic $y$-distortion of the neutrino background caused by scattering with energetic electrons or nucleons is {\it not} given by $\tilde{\Yspec}^\nu(x)=\Gspec^\nu\left[x\,\frac{\expf{x}-1}{\expf{x}+1}-4\right]$ as considered by \citep{Barenboim2025y}. This shape would appear only when taking the ensemble average of Fermi distributions with varying temperatures and equal weight, much like for the superposition of blackbody spectra \citep{Kamionkowski2003, Chluba2004, Stebbins2007}. The differences stem from the energy dependence of the scattering cross section and for the standard SZ effect the scattering process is equivalent to such an averaging process \citep{Zeldovich1972}.

\section{Single-scattering spectral evolution for thermal distributions}
\label{sec:sing_scat_col}
%-----------------------------------------------------
In this section, we illustrate the single-scattering limits of the evolution equations for various cases. This is essentially equivalent to evaluating the collision integral once for given incoming distributions.

%-----------------------------------------------------
\begin{figure}
    \centering
    \includegraphics[width=0.75\columnwidth]{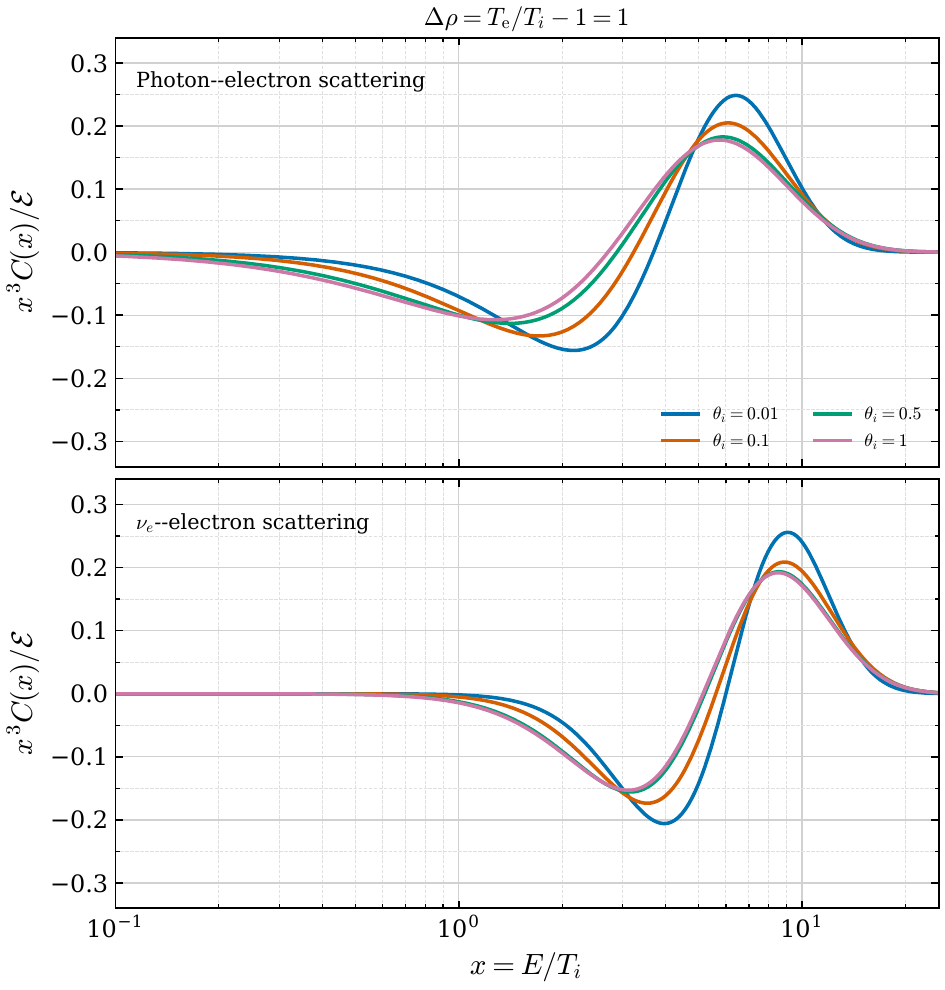}
    \vspace{-0mm}
    \caption{Collision term for photons (top) and neutrinos (bottom) scattering with electrons at varying temperatures $\theta_i$ ($i=\{\gamma, \nu\}$) and $\Delta \rho=1$. We scaled all cases by the energy exchange integral, $\mathcal{E}=\int x^3 \mathcal{C} \id x$, to normalize them. For increasing temperature, the importance of recoil again increases but less visibly, showing the increased importance of Doppler up-scattering terms.}
    \label{fig:Coll_Drho_1}
\end{figure}
%-----------------------------------------------------

%-----------------------------------------------------
\subsection{Varying temperature difference}
\label{sec:varying_DT}
%-----------------------------------------------------
We can now use Eq.~\eqref{eq:C_def_kernel_eval} to evaluate the collision term for thermal distributions of neutrinos and electrons (i.e., $\mu_i=0$) when the two distributions have a non-vanishing temperature difference. For comparison, we will also compute the same case for photon electron scattering. 

In Fig.~\ref{fig:Coll_Drho_0.001}, we show the collision term in the single-scattering evaluation for varying temperatures and small temperature difference with the electrons. We normalized all of the curves to carry the same total energy density. The collision term shows a characteristic sink-source structure from low to high energies as expected from up-scattering by electrons. With increasing temperature, the importance of recoil moves the null of the distribution to lower energies. This is consistent with the expectations from the single scattering energy exchange moments, which exhibit a similar behavior \citep{CSpack2019, Chluba2026nuKernels}. 
For photons, the single-scattering signal also extends to lower energies than for neutrinos, which is a reflection of the reduced scattering efficiency for neutrinos with energy, leading to stronger energy redistribution at high neutrino energies. In the figure, we used thermal distributions with $\hat{\mu}_i=0$, finding however that the evaluation does not depend much on this unless considering highly degenerate cases $\hat{\mu}_i>1$.

In Fig.~\ref{fig:Coll_Drho_1}, we show a similar sequence of collision term evaluations but for $\Te=2 T_\nu$ (i.e., $\rho=2$). In this case, the scattered distributions remain more focused towards the high energy part than in the previous case. Similarly, the shift of the distributions towards lower energies with increasing temperature is not as pronounced. Both aspects reflect the increased importance of Doppler terms in the redistribution process, causing enhanced up-scattering, as also expected from the scaling of the first energy moment of the kernels \citep{CSpack2019, Chluba2026nuKernels}.

%-----------------------------------------------------
\begin{figure}
    \centering
    \includegraphics[width=0.7\columnwidth]{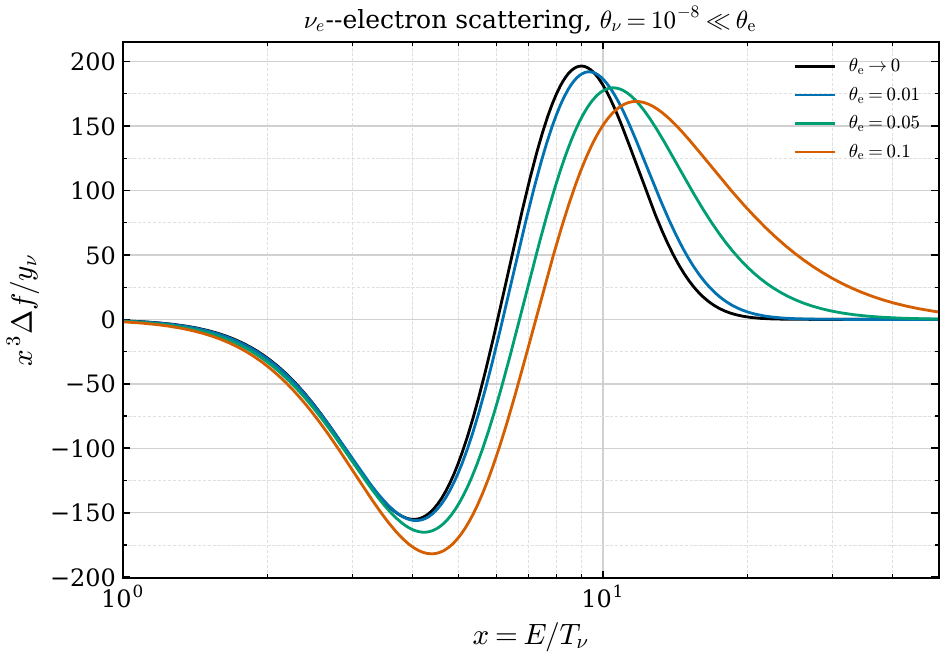}
    \vspace{-0mm}
    \caption{Illustration of the $\nu$SZ effect for varying temperatures of the electrons. We numerically evaluated the collision integral for $\theta_\nu\ll \The$. Like for the CMB SZ effect, the null of the distortion shifts towards higher frequencies, with the rough position scaling like $x_{\rm null}\approx 6.029+14.68\theta_{\rm e}-26.37\theta_{\rm e}^2
$ at $\The\lesssim 0.1$.}
    \label{fig:SZ-high-Coll}
\end{figure}
%-----------------------------------------------------

%-----------------------------------------------------
\subsection{SZ-like signal at high temperatures}
\label{sec:SZ_large}
%-----------------------------------------------------
As another illustration, we also evaluate the collision term in the $\nu$SZ limit, which assumes a single scattering distortion $(y_\nu \ll 1)$ and $T_\nu\ll \Te$. In Fig.~\ref{fig:SZ-high-Coll} we show the result of this computation. We used {\tt CSpack} for the computation but had to perform various Taylor series approximations to maintain numerical precision for the extreme temperature ratio. We have also confirmed this result to high-precision. In the future, using the Doppler-dominated limit of the scattering kernel like in the case for photons \citep{CSpack2019, Lee2024SZpack} will be beneficial.

We can see that the null of the $\nu$SZ distortion moves towards higher energies with increasing electron temperature (see caption of Fig.~\ref{fig:SZ-high-Coll} for approximate scaling), as is similar to the CMB relativistic SZ effect \citep[e.g.,][]{Itoh98}. 
We confirmed  that at $\The \lesssim 0.02$ the scattered signal is well-represented by the expansion, Eq.~\eqref{eq:Dn_nu}. However, even when we use a very high order expansion for the {\it formally exact} $\nu$SZ operator in Eq.~\eqref{eq:S_SZ_nu}, the approximation fails to converge for $\The \gtrsim 0.03$ and at $x\gtrsim 10$. 
We also note that for scattering nucleons, one can find $\theta_s\ll 1$ even when $\The\gg 1$ due to the extra mass suppression. The results in Fig.~\ref{fig:SZ-high-Coll} then carry over with appropriate redefinitions of $\The$ and $y_\nu$ as explained in section~\ref{sec:extensions}.

%-------------------------------------------------
\section{Numerical formulation and setups}
\label{sec:num_setup}
%-------------------------------------------------
In this section we describe the numerical setups for various scattering problems both in the Fokker-Planck type limit and using the scattering kernel treatment of the collision term. Much of the formalism was also described in \citep{Acharya2021, Acharya2022large}.

%-------------------------------------------------
\subsection{Kompaneets equation for photons with small distortions}
%-------------------------------------------------
For numerical applications we are interested in the case when the photon occupation number is given by $n=n_{\rm eq}(x)+\Delta n$ with blackbody spectrum $n_{\rm eq}(x)=1/(\expf{x}-1)$ and small distortion $\Delta n$.
Linearizing with respect to $\Delta n$, then yields
%-------------------------------------------------
\bealf{
\label{eq:gamma_Komp_linear}
\frac{\partial \Delta n}{\partial y}&\approx \frac{1}{x^2}\frac{\partial}{\partial x} x^4 \left[\frac{\partial \Delta n}{\partial x}+\Delta n(1+2 n_{\rm eq})\right]+Y \,\Delta \rho,
}
%-------------------------------------------------
with $\Delta \rho = \Te/\Tg-1$ and $y$-distortion $Y=\boostO(\boostO-3) n_{\rm eq}$. This equation can also be written as 
%-------------------------------------------------
\bealf{
\frac{\partial \Delta n}{\partial y}&\approx x^2 \Bigg[\Delta n'' 
+\left(\frac{4}{x}+A\right) \Delta n' 
+ \frac{4A-2G}{x} \Delta n \Bigg]  +Y \,\Delta \rho,
}
%-------------------------------------------------
with $A=1+2n_{\rm eq}$ and $G=\boostO n_{\rm eq}$. For the numerical solutions, we will assume that the electrons always take the Compton equilibrium temperature. At linear order, the relevant expression can again be obtained by integrating the evolution equation over $x^3 \id x$, setting the left hand side to zero and then solving for $\Delta \rho_{\rm C}$, yielding
%-------------------------------------------------
\bealf{
\label{eq:Drho_C}
\Delta \rho_{\rm C}&=-\frac{\int x^5 \left[\Delta n'' 
+\left(\frac{4}{x}+A\right) \Delta n' + \frac{4A-2G}{x} \Delta n\right]\!\id x}{\int x^3 Y(x) \id x}.
}
%-------------------------------------------------
For our evolution problems we will thus use $\Delta\rho \approx \Delta \rho_{\rm C}$. Although one could instead use the linearized version of Eq.~\eqref{eq:Te_C} \citep[see e.g., Eq.~(15) of][]{Chluba2011therm}, numerically Eq.~\eqref{eq:Drho_C} more consistently treats the related integrals within the chosen discretization scheme. The solutions of the problem are then set by the initial distortion and in the limit of many scatterings ($y\gg 1$), full equilibrium is reached. 

%-------------------------------------------------
\subsection{Kompaneets equation for neutrinos}
%-------------------------------------------------
Inserting $f=f_{\rm eq}(x)+\Delta f$ with $f_{\rm eq}(x)=1/(\expf{x}+1)$ and linearizing with respect to $\Delta f$, this then yields
%-------------------------------------------------
\bealf{
\label{eq:nu_Komp_linear}
\frac{\partial \Delta f}{\partial y}&\approx \frac{1}{x^2}\frac{\partial}{\partial x} x^6 \left[\frac{\partial \Delta f}{\partial x}+\Delta f(1-2 f_{\rm eq})\right]+Y^\nu \,\Delta \rho,
}
%-------------------------------------------------
with $\Delta \rho = \Te/T_\nu-1$ and $Y^\nu=\frac{1}{x^2}\frac{\partial}{\partial x} x^6 \frac{\partial}{\partial x}f_{\rm eq} =x^2 \boostO(\boostO-5) f_{\rm eq}$.
%-------------------------------------------------
This equation can also be written as 
%-------------------------------------------------
\bealf{
\frac{\partial \Delta f}{\partial y}&\approx x^4 \Bigg[\Delta f'' 
+\left(\frac{6}{x}+A^\nu\right) \Delta f' 
+ \frac{6A^\nu+2G^\nu}{x} \Delta f \Bigg]  +Y^\nu \,\Delta \rho,
}
%-------------------------------------------------
with $A^\nu=1-2f_{\rm eq}$ and $G^\nu=\boostO f_{\rm eq}$. This equation again has to be augmented with an equation to determine the electron temperature. As before, integrating the neutrino Kompaneets equation over $x^3 \id x$, setting the left hand side to zero and then solving for $\Delta \rho_{\rm C}^{\nu}$ yields
%-------------------------------------------------
\bealf{
\Delta \rho_{\rm C}^{\nu}&=-\frac{\int x^7 \left[\Delta f'' 
+\left(\frac{6}{x}+A^\nu\right) \Delta f' + \frac{6A^\nu+2G^\nu}{x} \Delta f\right]\!\id x}{\int x^3 {Y}^\nu(x) \id x}.
}
%-------------------------------------------------
We note, however, that in the presence of CMB photons the interactions with electrons are usually dominated by those, such that the electron temperature is close to Compton equilibrium with the photons. Below we will consider a range of situations for illustration.

%-----------------------------------------------------
\begin{figure}
    \centering
    \includegraphics[width=0.99\columnwidth]{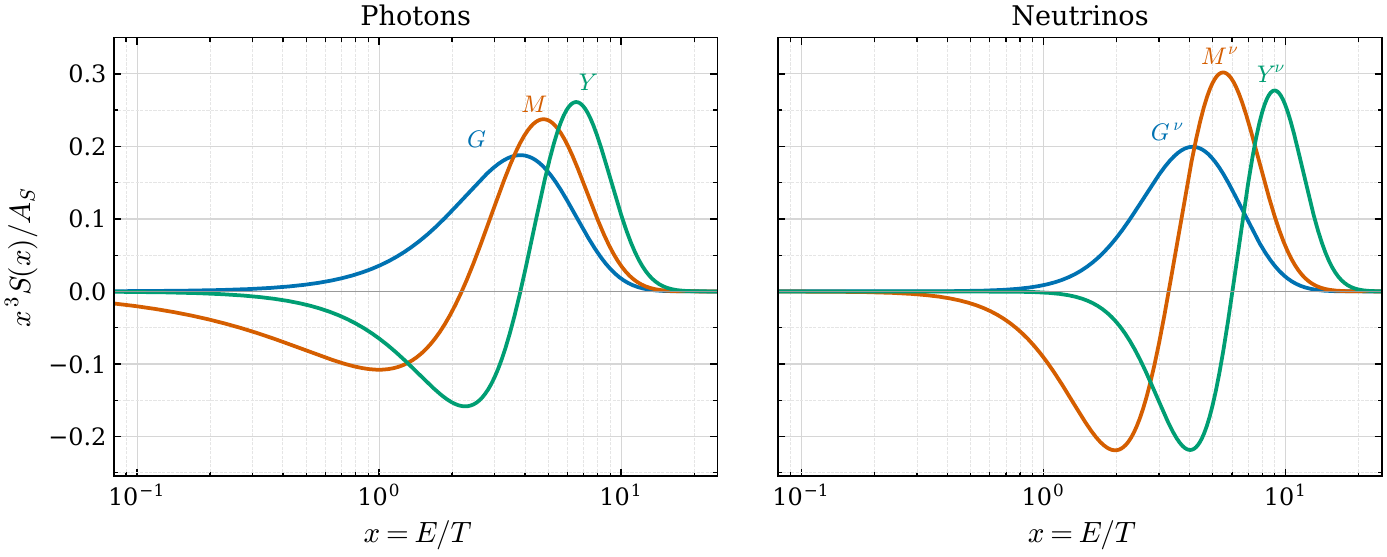}
    \vspace{-0mm}
    \caption{Main distortion shapes for photons (left) and neutrinos (right). All curves have been normalized to have unity energy density, $\int x^3 S/A_S \id x=1$, with normalization constants $\left\{A_G, A_M, A_Y\right\} \approx  \left\{25.98, 4.636, 25.98\right\}$ and $\left\{A_{G^\nu}, A_{M^\nu}, A_{Y^\nu}\right\} \approx  \left\{22.73, 1.502, 709.6\right\}$. }
    \label{fig:GMY}
\end{figure}
%-----------------------------------------------------

%-------------------------------------------------
\subsection{Main spectral shapes close to equilibrium}
\label{sec:M_distortion}
%-------------------------------------------------
In section~\ref{sec:nuSZ_basis} we discussed a spectral basis that spans all distortion shapes for photons and neutrinos introduced by boosts of the background equilibrium distribution. However, none of these functions describe the distribution that is reached asymptotically in the limit of many scatterings. Defining the photon Kompaneets operator, $\hat{\mathcal{K}}_x=\frac{1}{x^2}\frac{\partial}{\partial x} x^4 \left[\frac{\partial }{\partial x}+(1+2 n_{\rm eq})\right]$, it is easy to show that $\hat{\mathcal{K}}_x \frac{G}{x}=0$ and $\hat{\mathcal{K}}_x G =-Y$ \citep{chluba_spectro-spatial_2023-II, Chluba2026}. In a similar manner, we can then define the neutrino Kompaneets operator, $\hat{\mathcal{K}}^\nu_x=\frac{1}{x^2}\frac{\partial}{\partial x} x^6 \left[\frac{\partial }{\partial x}+(1-2 f_{\rm eq})\right]$, which obeys $\hat{\mathcal{K}}^\nu_x \frac{G^\nu}{x}=0$ and $\hat{\mathcal{K}}^\nu_x G^\nu =-Y^\nu$. With $\frac{G}{x}$ and $\frac{G^\nu}{x}$, we can then define the number-conserved distortion spectra
%-------------------------------------------------
\bealf{
M(x)&=G(x)\left[\beta_\gamma-\frac{1}{x}\right],
&
M^\nu(x)&=G^\nu(x)\left[\beta_\nu-\frac{1}{x}\right],
}
%-------------------------------------------------
with $\beta_{\gamma}=\frac{\pi^2}{18\zeta(3)}\approx 0.4561$ and $\beta_{\nu}=\frac{2}{3} \,\beta_{\gamma}=\frac{\pi^2}{27\zeta(3)}\approx 0.3041$. These are the $\mu$-distortion spectra for the photons and the neutrinos, with nulls at $x\approx 2.193$ and $x_\nu\approx 3.288$. 

The shapes of the main spectra are illustrated in Fig.~\ref{fig:GMY}, contrasting photon and neutrino distortions. This shows that for neutrinos, the distributions are generally shifted upward in energy. Importantly, in contrast to the photon case, the null of the $y$-type scattering distortion, $Y^\nu$, does not coincide with the maximum of the temperature shift spectrum, $G^\nu$, at $x_{\rm max}\approx 4.131$. This is due to the energy dependence of the scattering cross section.

Given the results above, it is then also easy to see that $\hat{\mathcal{K}}_x M=-\beta Y$ and $\hat{\mathcal{K}}^\nu_x M^\nu=-\beta_\nu Y^\nu$. Since we also have
%-------------------------------------------------
\bealf{
\label{eq:Drho_C_Cnu}
\Delta \rho_{\rm C}&=-\frac{\int x^3 \hat{\mathcal{K}}_x\Delta n \id x}{\int x^3 Y(x) \id x},
&
\Delta \rho^\nu_{\rm C}&=-\frac{\int x^3 \hat{\mathcal{K}}^\nu_x\Delta f \id x}{\int x^3 Y^\nu(x) \id x},
}
%-------------------------------------------------
we have $\Delta \rho_{\rm C}=\Theta+\beta \mu$ for $\Delta n= \Theta G+\mu M$ and similarly $\Delta \rho^\nu_{\rm C}=\Theta+\beta_\nu \mu$ for $\Delta f= \Theta G^\nu+\mu M^\nu$, with temperature parameter $\Theta$ and dimensionless chemical potential $\mu$. This implies that the solutions $\Delta n= \Theta G+\mu M$ and $\Delta f= \Theta G^\nu+\mu M^\nu$ remain unchanged when inserted into the photon and neutrino Kompaneets equations, Eq.~\eqref{eq:gamma_Komp_linear} and \eqref{eq:nu_Komp_linear}, respectively. They are {\it stationary solutions} in the small distortion limit and are also found asymptotically in the scattering kernel treatment.

%-------------------------------------------------
\subsection{Scattering kernel treatments}
%-------------------------------------------------
In Eq.~\eqref{eq:C_def_kernel_eval} we gave the full collision term for photons and neutrinos. We are now interested in the limit $\Delta n, \Delta f\ll 1$. For this we can generalize the setup for photons following \citep{Acharya2022large} but linearizing the final expression. Using $\Delta \phi=\phi-1\approx - \Delta \rho$, we then have
%-------------------------------------
\bsub
\begin{align}
\label{eq:C_def_kernel_eval_lin}
\mathcal{C}_{\pm}
&\approx
\int \id \omega' \, P(\omega \rightarrow \omega',\The, \mu_{\rm e})\times \mathcal{S}_\pm^{(1)}(\omega, \omega', \Delta \phi),
\\
\mathcal{S}_+^{(1)}(\omega, \omega', \Delta \phi)&=
\expf{x-x'}f^+_{xx'}\,\Delta f(x)-f^+_{x'x}\,\Delta f(x')+(x-x')\,\Delta \phi\,f^+_{x'x}\,\frac{G^\nu(x')}{x'}
\\
\mathcal{S}_-^{(1)}(\omega, \omega', \Delta \phi)&=
\expf{x-x'}f^-_{xx'}\,\Delta n(x)-f^-_{x'x}\,\Delta n(x')+(x-x')\,\Delta \phi\,f^-_{x'x}\,\frac{G(x')}{x'}
\end{align}
\esub
%-------------------------------------
with $f^\pm_{xx'}=(1\pm\expf{-x})/(1\pm\expf{-x'})$, which follows from $f^-_{xx'}=[1+n_{\rm eq}(x')]/[1+n_{\rm eq}(x)]$ for photons and $f^+_{xx'}=[1-f_{\rm eq}(x')]/[1-f_{\rm eq}(x)]$ for neutrinos, and $x=\omega/T_{\rm r}$, $x'=\omega'/T_{\rm r}$. As expected, the scattering interactions cause a redistribution between the populations in the frequency bins $x\leftrightarrow x'$. In addition, a source term that is proportional to $\Delta \phi$ appears, which at low temperatures sources a $y$-type distortion. Both terms can be modeled as a scattering matrix multiplication in evaluations of the collision term \citep{Acharya2022large}. To compute the scattering matrix, we follow the procedure as described in sections~2.2.3 and 2.2.4 of \citep{Acharya2022large} for the linearised equations.

%-----------------------------------------------------
\begin{figure}
    \centering
    \includegraphics[width=0.99\columnwidth]{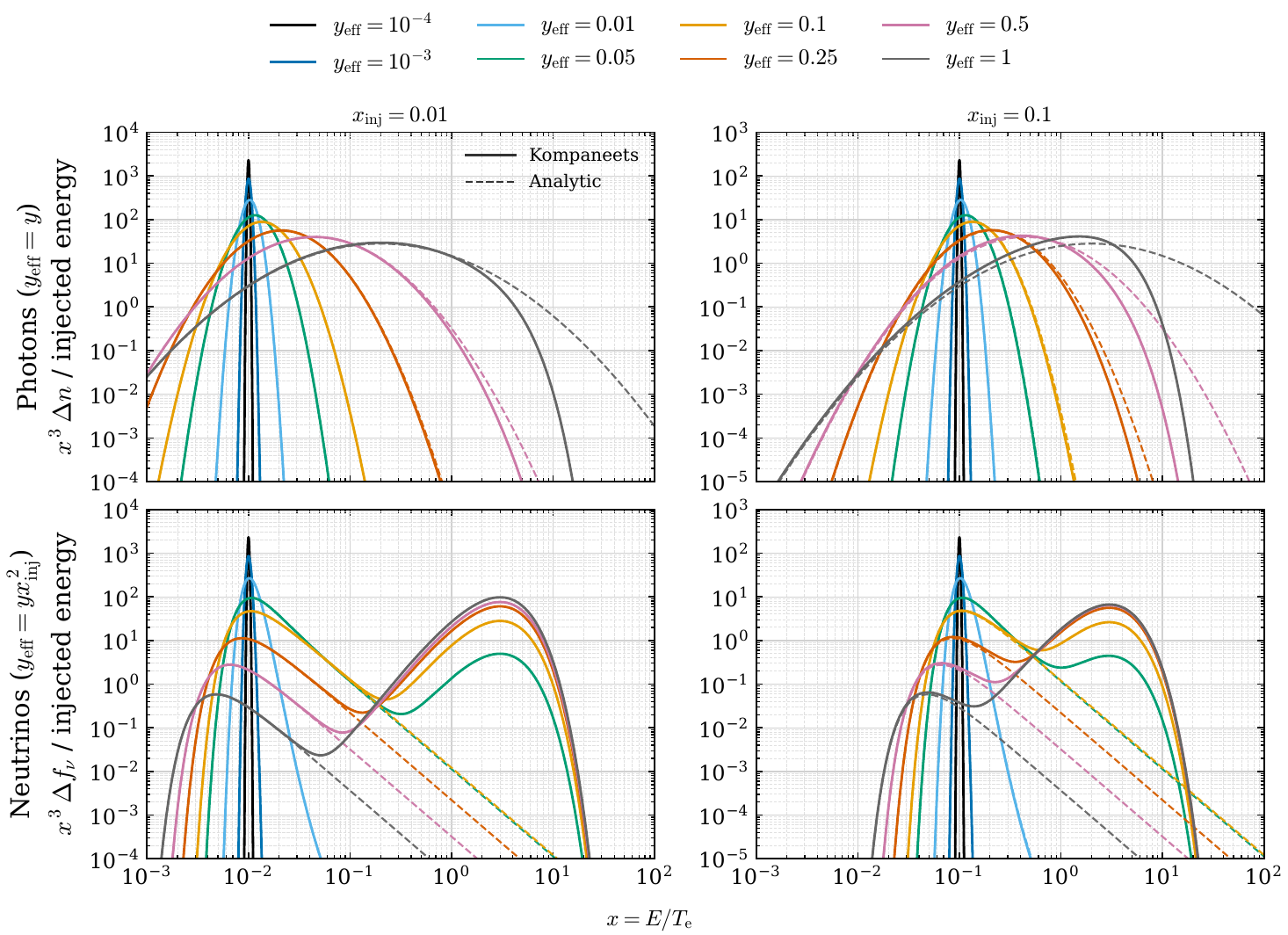}
    \vspace{-0mm}
    \caption{Line injection solutions for photons (upper panels) and neutrinos (lower panels) at $x_{\rm inj}=10^{-2}$ and $x_{\rm inj}=10^{-1}$. We compare the solutions obtained with the Kompaneets type equations to the analytic approximations in the Doppler-dominated scattering regime (see section~\ref{sec:Doppler_dom_sol}). For the neutrino case, the effective scattering $y$-parameter was set to $y_{\rm eff}=y x_{\rm inj}^2$, such that the runtime increases with decreasing $x_{\rm inj}$. The analytic solutions work well at small scattering $y$-parameter but fail strongly at high frequencies and $y_{\rm eff}\gtrsim 0.5$.}
    \label{fig:line_evo_Kompaneets_I}
\end{figure}
%-----------------------------------------------------

%-------------------------------------------------
\subsection{Solver details}
%-------------------------------------------------
To solve the Kompaneets-type equations, we use a 5-point stencil based on Lagrange polynomials to discretize the frequency grid \citep{Acharya2022large}. Uniform and non-uniform frequency grids are allowed and all functions relating to the background spectrum (e.g., $A$, $G$ and $Y$) are precomputed once the grid is set. The stiff system of coupled ordinary differential equations (ODEs) is then solved using an implicit Gears method with variable time-step \citep[see][for details]{Chluba2010}. We found a fourth order implicit time-stepper to be most efficient and the banded matrix is solved using sparse Gaussian elimination, with sparse matrix storage whenever possible. Overshooting is generally permitted, although for long evolution sequences it is kept more limited.

For the scattering kernel treatment, we use the efficient setup routines of {\tt CSpack} \citep{CSpack2019, Acharya2022large}, which have now been generalized to also allow the modeling of neutrino kernels \citep{Chluba2026nuKernels}. In particular, we use interpolated thermally-averaged scattering kernels to accelerate the setup of the scattering matrices. Convergence of the results with respect to various parameter choices such as the wing truncation and grid densities were confirmed. Detailed balance properties and Fermi-blocking effects are directly worked into the setup to ensure that the correct equilibrium distribution is approached \citep{Acharya2022large}. The same ODE solver as for the FP type problems was used but with a second order implicit time-stepper and a bi-conjugate gradient method to solve the dense matrix problems.

For both types of problems, we use a combination of absolute and relative precision requirements, aiming at an overall precision of $0.1\%$ for the spectrum. For FP type problems, this implies $\simeq 2000-5000$ frequency grid points while for the kernel treatments one can get away with $\simeq 300-800$ grid points. For line injection problems we sample the region around the injection point more finely to improve precision at the initial stages of the run. 
Conservation of photon and neutrino number as well as energy conservations were monitored and found to be sufficiently stable. Convergence tests with solver settings were passed. 

%-----------------------------------------------------
\begin{figure}
    \centering
    \includegraphics[width=0.95\columnwidth]{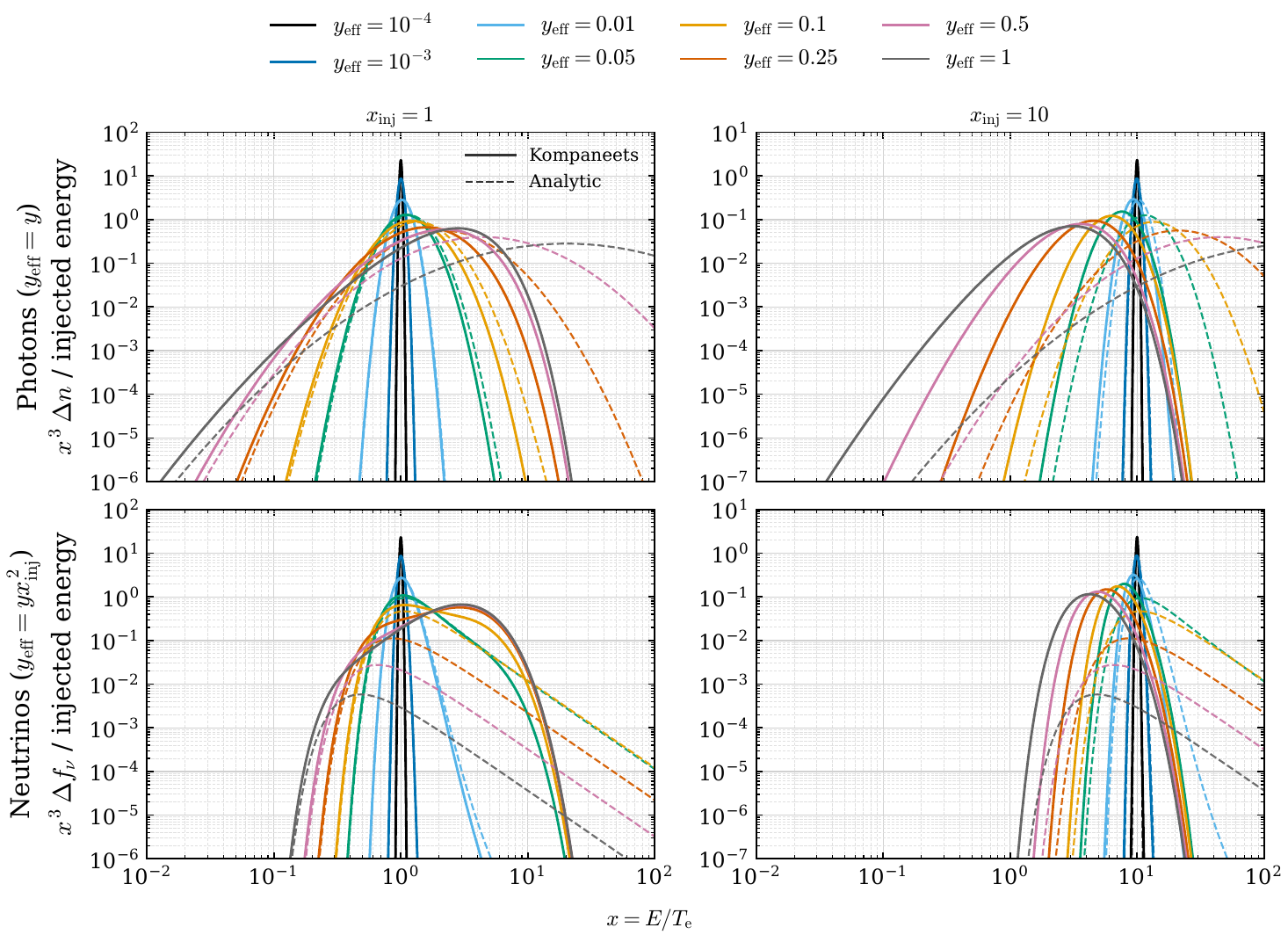}
    \vspace{-0mm}
    \caption{Same as for Fig.~\ref{fig:line_evo_Kompaneets_I} but for $x_{\rm inj}=1$ and $x_{\rm inj}=10$. The analytic solutions do not capture the evolution well especially at $x\gtrsim 1$ since recoil is not accounted for.}
    \label{fig:line_evo_Kompaneets_II}
\end{figure}
%-----------------------------------------------------

%-------------------------------------------------
\section{Numerical treatment of repeated scattering problems}
\label{sec:num_sols}
%-------------------------------------------------
We are now interested in studying numerical solutions to scattering problems. We will first consider the modeling using the Fokker--Planck approximation, which formally is only valid in the low-temperature limit. However, as we shall see, it can also be used in the high temperature regime to capture the overall dynamics of the solution and gain physical intuition. We then continue by solving the exact scattering integral using the methods developed for the repeated scattering of photons by free electrons \citep{Acharya2021, Acharya2022large}. For comparison, we also present the scattering of photons, which allows us to illustrate similarities and differences in the physics for neutrinos.

%-----------------------------------------------------
\begin{figure}
    \centering
    \includegraphics[width=0.95\columnwidth]{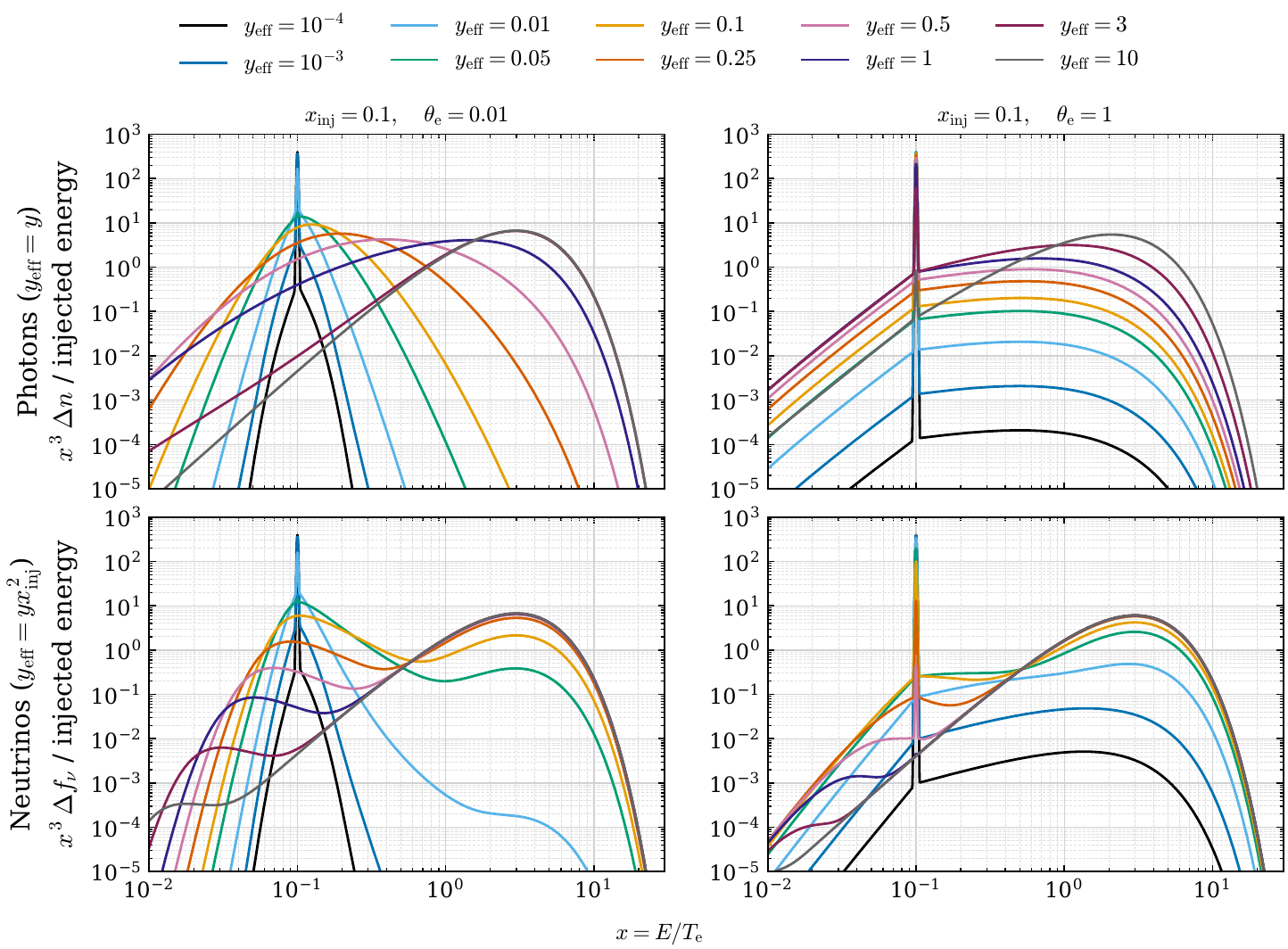}
    \vspace{-0mm}
    \caption{Line injection solutions for photons (upper panels) and neutrinos (lower panels) at $x_{\rm inj}=0.1$. Here we used the scattering kernel setup for $\The=0.01$ (left column) and $\The=1$ (right column) to solve the problem. For the neutrino case, the effective scattering $y$-parameter was set to $y_{\rm eff}=y x_{\rm inj}^2$.}
    \label{fig:line_evo_Kernel_I}
\end{figure}
%-----------------------------------------------------

\begin{comment}
%-----------------------------------------------------
\begin{figure}
    \centering
    \includegraphics[width=1\columnwidth]{Figures/kernel_line_evolution_xinj_1_theta_comparison.pdf}
    \vspace{-0mm}
    \caption{Same as for Fig.~\ref{fig:line_evo_Kernel_I} but for $x_{\rm inj}=1$.}
    \label{fig:line_evo_Kernel_II}
\end{figure}
%-----------------------------------------------------
\end{comment}

%-----------------------------------------------------
\begin{figure}
    \centering
\includegraphics[width=0.92\columnwidth]{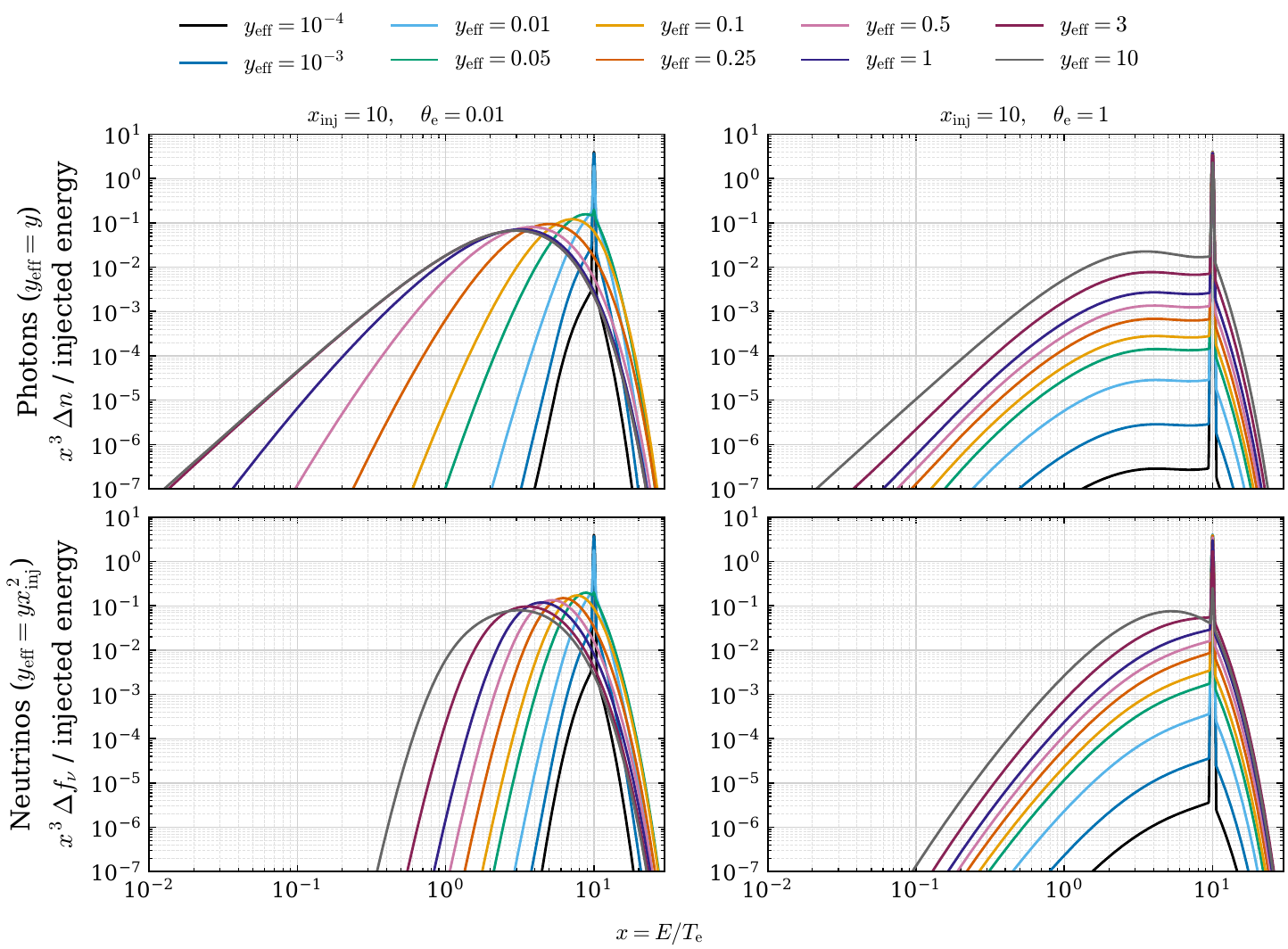}
    \vspace{-0mm}
    \caption{Same as for Fig.~\ref{fig:line_evo_Kernel_I} but for $x_{\rm inj}=10$.}
    \label{fig:line_evo_Kernel_III}
\end{figure}
%-----------------------------------------------------

\subsection{Line injection problems}
\label{sec:line_inj}
%-----------------------------------------------------
For illustration we first consider a few single line injection problems. We start with a narrow Gaussian in energy to model a $\delta$-function at $x_{\rm inj}=0.01, 0.1, 1$ and $10$ with the FP treatment for illustration. We omit the effect of background equilibrium distributions for now assume that the scattering electrons have a fixed temperature.
The results are summarised in Fig.~\ref{fig:line_evo_Kompaneets_I} and \ref{fig:line_evo_Kompaneets_II}. For comparison, we also show the analytic solution based on Doppler-dominated scattering (see section~\ref{sec:Doppler_dom_sol}) which neglects the effect of recoil and is thus not expected to work well for large scattering $y$-parameter. 

Focusing on Fig.~\ref{fig:line_evo_Kompaneets_I}, we can see that the overall evolution is dominated by an up-scattering of the initial line by the thermal electrons. For the photons, the distribution remains rather symmetric until late phases of the evolution, when recoil terms prevent photons from further up-scattering in the high energy tail. This aspect is clearly not captured by the analytic approximation, which develops a heavy tail towards large $x$ for $y\gtrsim 1$. 

For the neutrino cases, we scaled the scattering $y$-parameter to yield a constant effective parameter $y_{\rm eff}=y x_{\rm inj}^2$. This is because at low energies, the scattering efficiency reduces as $x^2$ due to the energy-dependence of the scattering cross section, meaning that the evolution of the distribution slows as $x\rightarrow 0$. One can clearly see a significant asymmetry in the scattering process, with slower evolution towards lower energies (i.e., less drifting and broadening) than towards higher energies. 
The latter is strikingly visible for the case $x_{\rm inj}=0.01$, where the neutrino distribution reaches a quasi-equilibrium at high frequencies while the photon distribution continues to evolve for $y_{\rm eff}\rightarrow 1$.

In contrast, the analytic approximation significantly overestimates the efficiency of up-scattering, creating a power-law high-energy spectrum, far outside the computational window. Indeed, neutrinos leave the computational domain with the analytic approximation due to the lack of recoil, which for the numerical FP solution causes a pile-up of neutrinos in a Wien-type $\Delta f \simeq \expf{-x}$ distribution.

A very similar qualitative picture presents itself for injections at $x_{\rm inj}=1$ and $10$ (see Fig.~\ref{fig:line_evo_Kompaneets_II}). The stark difference is that in the FP treatment the evolution of the distributions towards low energies is now more pronounced, especially for $x_{\rm inj}=10$. The analytic approximations again strongly overshoot towards high energies and also do not show as strong a drift towards lower energies due to the absence of recoil contributions. When compared to the photon case, the neutrino solutions again display a clear slowing of the redistribution process towards lower energies, simply due to the $\simeq x^2$-scaling of the scattering efficiency. 

We also stress that the solution only depends on $x$ and the $y$-parameter, but not explicitly on the assumed temperature of the electrons. This aspect changes when using the exact scattering kernel treatment, as we illustrate next. In addition, analytic approximations based on pure recoil, Eq.~\eqref{eq:sol_recoil_only}, fail to capture the clearly visible broadening of the numerical solution.

In Fig.~\ref{fig:line_evo_Kernel_I} and \ref{fig:line_evo_Kernel_III} we show a similar evolution sequence as for the FP solutions but this time using the linearized kernel treatment. In each figure, we fix $x_{\rm inj}$ and compare a low and high temperature run for photons and neutrinos. The key features of the FP treatment are reproduced, but especially early on, while the distribution is still very narrow, the kernel solutions exhibit more structure that reflects the underlying scattering dynamics. The higher temperature cases show a much more broad redistribution function, with distant jumps out of the initial injection line core. Neutrino scattering is less efficient at lower energies showing that convergence towards a Wien spectrum is inhibited, while equilibrium is reached rather quickly at high energies, once some particles populate this energy domain. For photons, the evolution tends towards equilibrium, with less dependence on the energy.

Figure~\ref{fig:line_evo_Kernel_III} illustrates the significant asymmetry of the scattering kernel at high energies, with a heavier red-scattering wing that is clearly visible for the high temperature case. Again the drifts towards low energies are significantly slower for the neutrino case than for the photons, as expected from the energy dependence of the scattering efficiency. This also means that at $y_{\rm eff}=10$, the neutrino distribution has not yet fully relaxed to a stationary state. This is also found in the FP solutions of the problem, but without capturing the full temperature dependence of the redistribution kernel.

%-----------------------------------------------------
\begin{figure}
    \centering
\includegraphics[width=0.8\columnwidth]{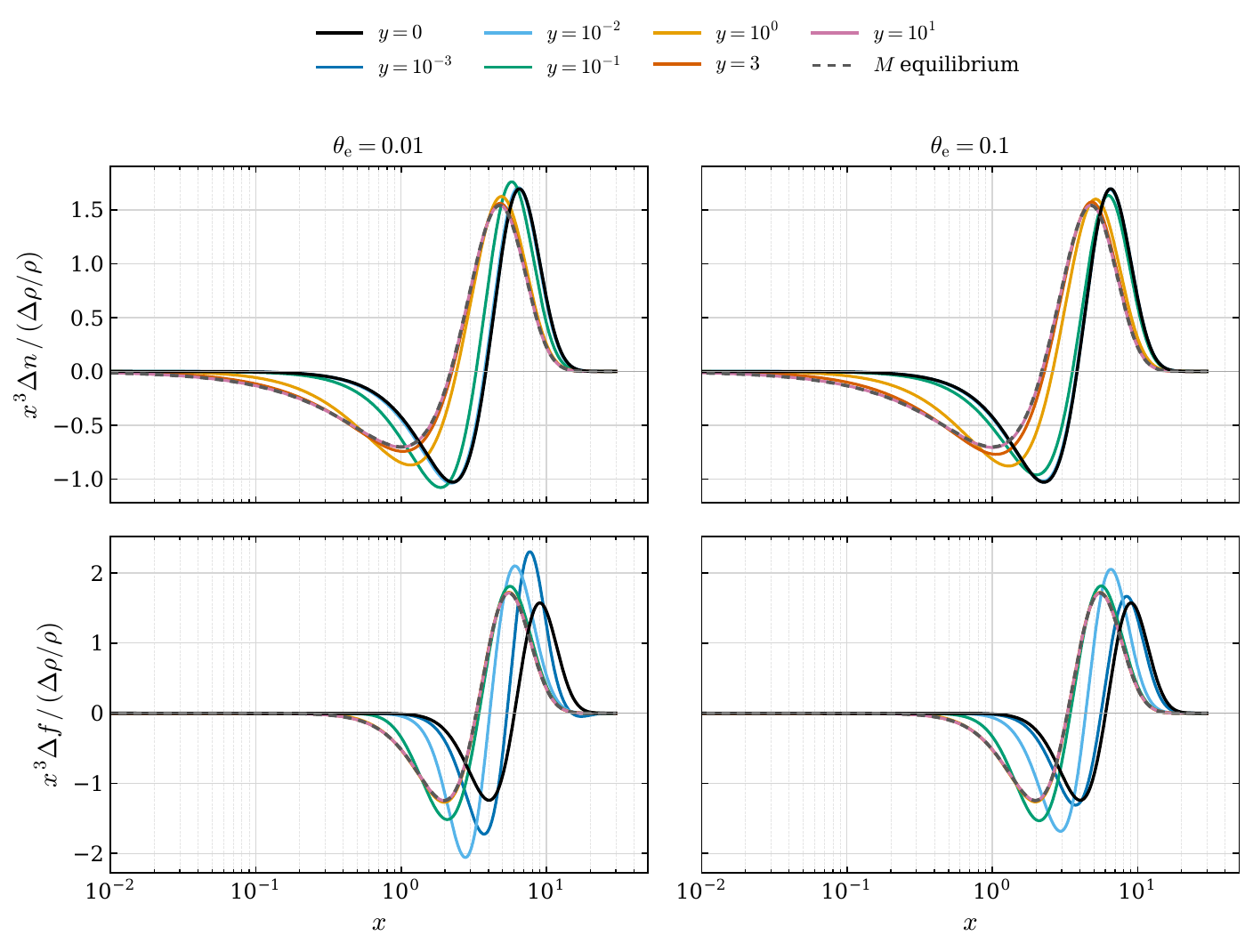}
    \vspace{-0mm}
    \caption{Snapshots for the evolution of the distributions towards kinetic equilibrium with the electrons after a single initial energy injection. The ambient temperature of the medium is varied to highlight the thermalization efficiency on the typical energy of the scattering particles.}
    \label{fig:thermal_evo_Kernel_I}
\end{figure}
%-----------------------------------------------------

%-----------------------------------------------------
\begin{figure}
    \centering
\includegraphics[width=0.8\columnwidth]{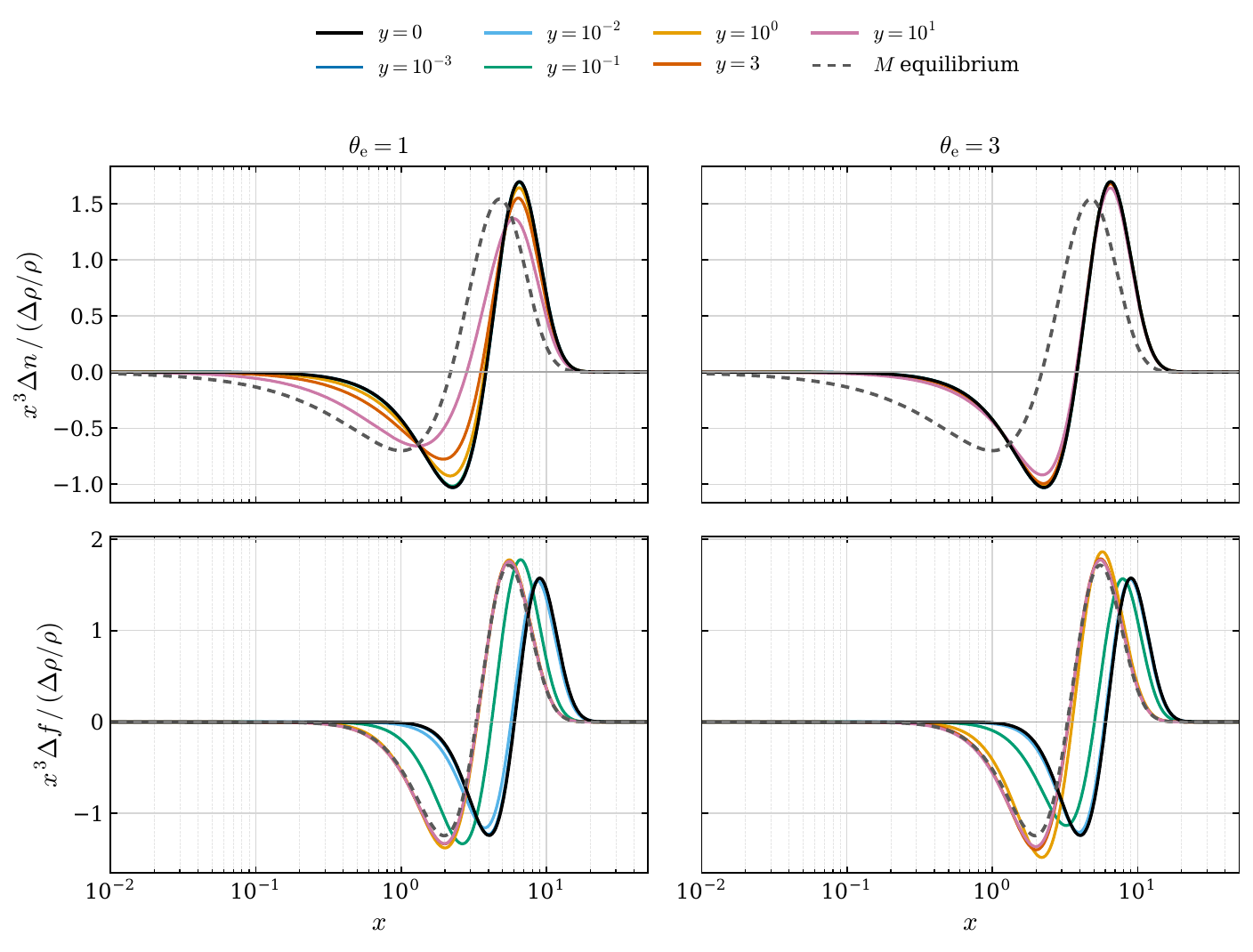}
    \vspace{-0mm}
    \caption{Same as Fig.~\ref{fig:thermal_evo_Kernel_I} but for $\The=1$ and $3$. The evolution is slower at higher temperatures.}
    \label{fig:thermal_evo_Kernel_II}
\end{figure}
%-----------------------------------------------------

\subsection{Single energy injection into a thermal background}
\label{sec:Y_inj}
%-----------------------------------------------------
We next consider examples where we imagine a single heating of the electrons that then quickly causes a $y$-type distortion which gradually transforms into a chemical potential distortion. This process is well studied in connection with the thermalization of CMB spectral distortions \citep{Chluba2011therm, Khatri2012mix, Chluba2013Green}, and here is illustrated at higher temperatures for photons and neutrinos. 

For simplicity, we shall assume that in all cases, the initial distortion is given by $Y$ and $Y^\nu$ with a fixed small energy density $\Delta \rho/\rho\simeq 10^{-5}$ to avoid and non-linear effects. At each timestep, we set the electron temperature to the equilibrium temperature in the distorted photon / neutrino field.

In Figs.~\ref{fig:thermal_evo_Kernel_I} and \ref{fig:thermal_evo_Kernel_II} we show snapshots of the evolution towards the stationary solution (see section~\ref{sec:M_distortion}) for varying background temperatures. For the cases shown in Fig.~\ref{fig:thermal_evo_Kernel_I}, kinetic equilibrium with the scattering electrons is indeed reached by $y=10$, however, for higher temperature media the distribution remains out of equilibrium (see Fig.~\ref{fig:thermal_evo_Kernel_II}). For both the photons and neutrinos, this is due to the significant dependence of the scattering probability on the temperature \citep{CSpack2019}. 

For the neutrinos, one also observes a differential transport time-scale, with the low energy part of the distribution relaxing less rapidly. This is again in line with the discussions in the previous section and stems from the $E_\nu^2$ scaling of the scattering cross section. 
We also remark that the equilibrium distribution is  independent of the temperature of the medium. This is because we are in the regime of small distortions as is familiar from photon scattering problems \citep[e.g.,][]{Chluba2020large}. 
We also confirmed that adding final state Pauli-blocking for the electrons causes percent-level changes to the curves.

%-----------------------------------------------------
\section{Efficiency of neutrino-$e^\pm$ scattering in the early Universe}
\label{sec:y_sc}
%-----------------------------------------------------
To illustrate when some of the solutions we presented might be relevant, we ask the question at which moment during cosmic history does the neutrino distribution become frozen with respect to electron scattering. 
For photon scattering, this occurs around a redshift $z\simeq \pot{5}{4}$ \citep[e.g.,][]{Hu1993}, however, for neutrinos the picture is a bit more involved. Firstly, due to the $\propto \bar{\sigma}_1 \omega_1^2$ scaling of the electron scattering cross section, the moment at which the neutrino distribution freezes depends on energy, an effect that we have already discussed above. In addition, in the cosmic plasma, the freezing occurs during and before electron-positron annihilation, because otherwise the electron densities are far too small to make a significant differences around the main part of the distribution. 

To calculate the various regimes, we can now define the look-back optical depth and (leading order Doppler-dominated) scattering $y$-parameter
%-------------------------------------
\bsub
\label{eq:tau_y}
\begin{align}
\tau_{\rm sc}&=\int_{t_i}^{t_0} \Ne \bar{\sigma}_1 \omega_1^2 \id t\approx 
\bar{\sigma}_1 x_1^2 \int_{T_0}^T \, \Ne(T)\,\frac{T^2}{\me^2} \,\frac{\id T}{T H}
\\
y_{\rm sc}&=\int_{t_i}^{t_0} 6 \The \Ne \bar{\sigma}_1 \omega_1^2 \id t
\approx 
\bar{\sigma}_1 x_1^2 \int_{T_0}^T 6\, \Ne(T)\,\frac{T^3}{\me^3} \,\frac{\id T}{T H},
\end{align}
\esub
%-------------------------------------
where we assume one common temperature, $T$, for all interacting species. In the above, we have taken the low energy limit of the following more general expressions\footnote{Here we neglect a small correction due to the entropy production and related departure from the $T\propto a^{-1}$ scaling.} 
%-------------------------------------
\bsub
\label{eq:tau_y_moments}
\begin{align}
\tau^{*}_{\rm sc}&\approx 
\bar{\sigma}_1 x_1^2 \int_{T_i}^T \frac{\Ne(T)\,\Sigma_0(\omega_1, T)}{\bar{\sigma}_1\omega_1^2}\,\frac{T^2}{\me^2} \,\frac{\id T}{T H}, 
\\
y^{*}_{\rm sc}&\approx 
\bar{\sigma}_1 x_1^2 \int_{T_i}^T \frac{\Ne(T)\,\Sigma_1(\omega_1, T)}{\bar{\sigma}_1\omega_1^2}\,\frac{T^2}{\me^2} \,\frac{\id T}{T H}
\end{align}
\esub
%-------------------------------------
to account for changes to the scattering efficiency as a function of the temperature and neutrino energy. In this expression, $\Sigma_k(\omega_1, T)$ is the thermally-averaged moment of the scattering kernel \citep{Chluba2026nuKernels}, which is defined in a way similar to the scattering of photons by electrons \citep{CSpack2019, Acharya2021}
%-------------------------------------
\begin{align}
\Sigma_k(\omega_1, T, \mu_{\rm e})
&=\int \id \omega_3 \,
\left(\frac{\omega_3-\omega_1}{\omega_1}\right)^k \, P(\omega_1 \rightarrow \omega_3, T/\me, \mu_{\rm e}) \, [1-f(\omega_3)].
\end{align}
%-------------------------------------
For completeness we allowed for degeneracy effects in the electron distribution through $\mu_{\rm e}$. However, close to equilibrium ($\mu_{\rm e}\approx 0$), these effects usually remain small (i.e., a few percent).

\paragraph{Post-annihilation era.}
We can now easily show that $\tau^{*}_{\rm sc}$ and $y^{*}_{\rm sc}$ are always much smaller than unity at energies where most of the neutrinos are in the distribution (i.e., $x \simeq 3.151$ for a relativistic Fermi-Dirac distribution) and temperatures $T\leq 20 \,{\rm keV}$ $(\theta\leq 0.04)$ when the freeze-out abundance is set (i.e., $z\ll 10^7-10^8$). To demonstrate this we use the free electron number density
%-------------------------------------
\begin{align}
\label{eq:Ne_freeze}
\Ne^{\rm f}\approx (1-Y_{\rm p}/2) \, N_{\rm b} = (1-Y_{\rm p}/2) \,\eta_{\rm b} \,N_{\gamma}\approx \pot{5.4}{-10}\,\left[\frac{1-Y_{\rm p}/2}{0.88}\right]\left[\frac{\eta_{\rm b}}{\pot{6.1}{-10}}\right] N_{\gamma} ,
\end{align}
%-------------------------------------
before recombination with helium mass fraction, $Y_{\rm p}\approx 0.24$, baryon-to-photon ratio $ \eta_{\rm b}$ and photon number density $N_{\gamma} \approx \frac{2\zeta(3)}{\pi^2}\,T^3$. Since $\theta \lesssim 0.04$, we may approximate $\tau^*_{\rm sc}\approx \tau_{\rm sc}$ and $y^*_{\rm sc}\approx y_{\rm sc}$. With $H\approx 1.66 \sqrt{g_*} \,\frac{T^2}{m_{\rm pl}}$ and Eq.~\eqref{eq:tau_y} we then find
%-------------------------------------
\bsub
\label{eq:tau_y_est}
\begin{align}
\tau_{\rm sc}&\approx 
\frac{2\zeta(3)}{1.66 \,\pi^2} \,
\frac{Y_{\rm e} \,\eta_{\rm b}\, \bar{\sigma}_1 x_1^2\,m_{\rm pl}\,\me}{\sqrt{g_*}} \,
\int_{\theta_0}^\theta \,\theta^2 \,\id \theta\approx
\frac{2\zeta(3)}{1.66 \,\pi^2} \,
\frac{Y_{\rm e} \,\eta_{\rm b}\, \bar{\sigma}_1 x_1^2\,m_{\rm pl}\,\me}{3\sqrt{g_*}} \,
\theta^3 
\\
y_{\rm sc}&\approx 
\frac{12\zeta(3)}{1.66 \,\pi^2}\,
\frac{Y_{\rm e} \,\eta_{\rm b}\, \bar{\sigma}_1 x_1^2\,m_{\rm pl}\,\me}{\sqrt{g_*}}\,
\int_{\theta_0}^\theta \,\theta^3 \,\id \theta
\approx 
\frac{12\zeta(3)}{1.66 \,\pi^2}\,
\frac{Y_{\rm e} \,\eta_{\rm b}\, \bar{\sigma}_1 x_1^2\,m_{\rm pl}\,\me}{4\sqrt{g_*}}\,
\theta^4
\approx
\frac{9}{2} \theta \,\tau_{\rm sc}(x_1, \theta),
\end{align}
\esub
%-------------------------------------
with electron mass fraction $Y_{\rm e}=1-Y_{\rm p}/2$, $\theta=T/\me$ and effective relativistic spin degrees of freedom, $g_*(T)$. 
Plugging in numbers, we then obtain
%-------------------------------------
\begin{align}
\label{eq:tau_y_est_eval}
\tau_{\rm sc}(x_1, \theta)&\approx
\pot{1.69}{-12}\,x_1^2\,\theta^3\,\left[\frac{Y_{\rm e}}{0.88}\right]\left[\frac{\eta_{\rm b}}{\pot{6.1}{-10}}\right]
\left[\frac{\bar{\sigma}_1}{\pot{7.38}{-45}\,\cm^2}\right]
\left[\frac{g_*}{3.38}\right]^{-1/2}, 
\end{align}
%-------------------------------------
where we used $\bar{\sigma}_1\approx \pot{1.11}{-20}\sigT$ for $\nu_e-e$-scattering. 
This means that at $\theta \simeq 0.04$ and $x_1\lesssim 10$ the scattering optical depth (which is a proxy for the rate of isotropization by electron scattering) is already far from unity. Similarly, energy redistribution is very inefficient.
The reason is of course that after the electron-positron annihilation era the number density of electrons and positron is on average too small to have a significant effect.
The simplistic estimate changes when we include the increased  scattering efficiency using the moments, but the final conclusion will remain the same.

%-----------------------------------------------------
\begin{figure}
    \centering
    \includegraphics[width=0.94\columnwidth]{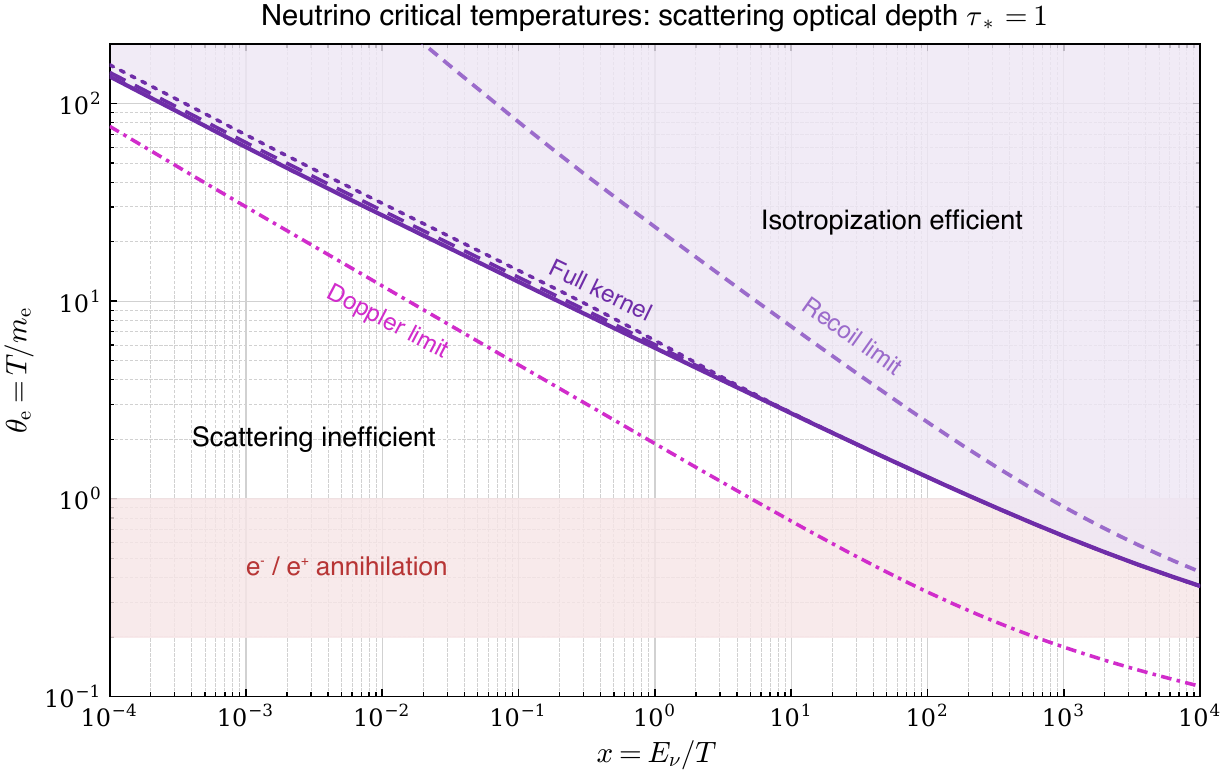}
    \\[3mm]
    \includegraphics[width=0.94\columnwidth]{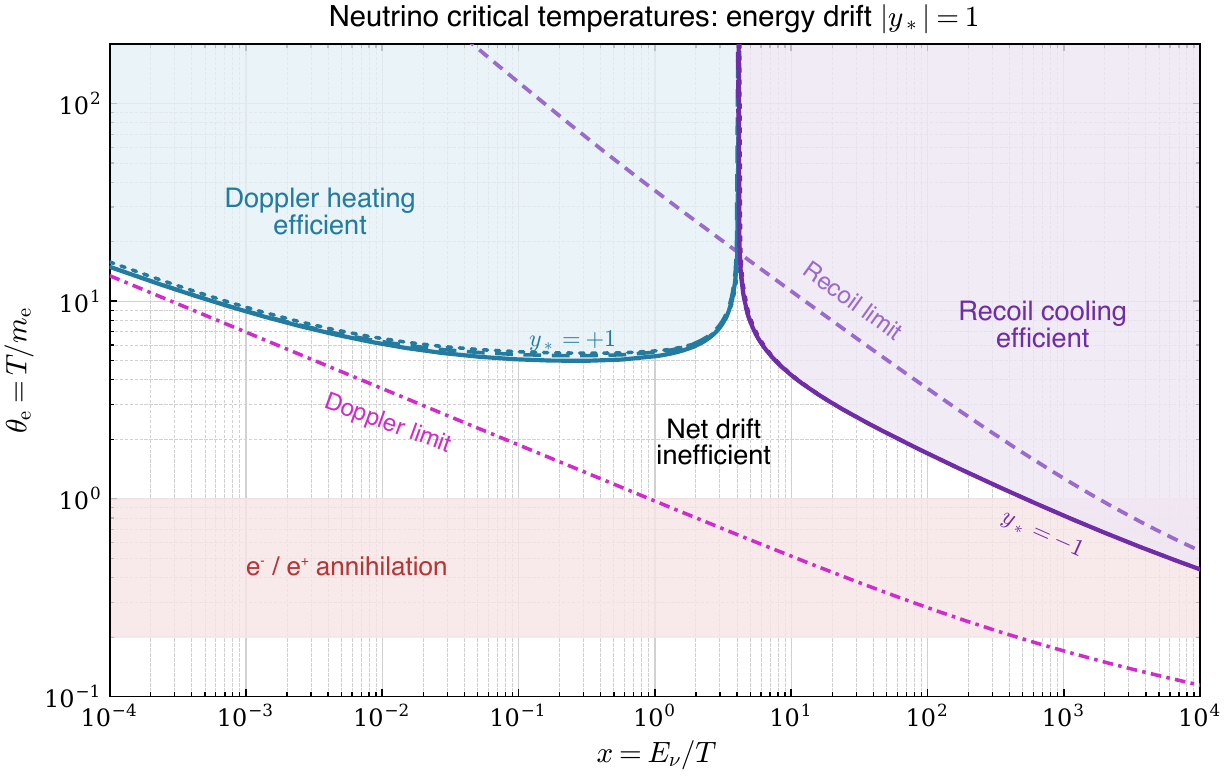}
    \vspace{-0mm}
    \caption{Critical temperatures for $\nu_e- e^\pm$-scattering. The upper panel illustrates the rate of isotropization using the zeroth moment of the scattering kernel ($\tau^*_{\rm sc}=1$), while the lower panel shows the domains of efficient energy exchange estimates through $|y^*_{\rm sc}|=1$. In both cases, we also show the estimated domain boundaries using the approximate moment expressions in the Doppler- or recoil-dominated regime. We also marked the temperature region when most of the $e^{\pm}$ pairs source photons and neutrinos. The extra lines close to the 'Full kernel' domain boundaries illustrate final state blocking effects (see text for details).}
    \label{fig:critical_domains_Sigma_0}
\end{figure}
%-----------------------------------------------------

\paragraph{Pre-annihilation era.}
We now assume that at $x_1\lesssim 100$, we will always find a critical temperature $\theta >0.2$ when we are still far from electron freeze-out as set by the baryon asymmetry. We can therefore assume equilibrium distributions for the electrons and positrons with $\mu_i=0$
%-------------------------------------
\begin{align}
\label{eq:Ne_Np_equ}
N^{\rm eq}_\pm \approx \frac{\me^3}{\pi^2}\int \frac{p^2 \id p}{\expf{\gamma/\theta}+1}
%=\frac{1}{\pi^2}\int_{1}^\infty \frac{\gamma \sqrt{\gamma^2-1} \id \gamma}{\expf{\gamma/\theta}+1}
%=\frac{\theta^3}{\pi^2}\int_{1/\theta}^\infty \frac{y \sqrt{y^2-1/\theta^2}\id y}{\expf{y}+1},
=\frac{T^3}{\pi^2}\int_0^\infty \frac{y^2 \id y}{\expf{\sqrt{1+\theta^2 y^2}/\theta}+1}
\approx \frac{3\zeta(3)}{2\pi^2}\,T^3\times \frac{K_2(1/\theta)}{2\theta^2},
\end{align}
%-------------------------------------
where we introduced $y=p/\theta$. The last expression gives a very good overall approximation at $\theta>0.1$. We note that $\frac{K_2(1/\theta)}{2\theta^2}\rightarrow 1$ for $\theta\gg 1$ giving the ultra-relativistic limit for $N_\pm$. 

We now compute the results numerically. We first implement the thermally-averaged normalized moments, $\hat{\Sigma}_k(x_1, \theta)=\Sigma_k(\omega_1=\theta\,x_1, \theta)/[\bar{\sigma}_1\omega_1^2]$ for $k=0$ and $k=1$ for both $\nu_e-e^-$ and $\nu_e-e^+$ scattering using {\tt CSpack}. For the moments, we do not include any final-state Fermi-blocking unless stated otherwise. To improve the numerical stability, we use the exact analytic moment expressions from section~2.2 of \citep{Chluba2026nuKernels} at low neutrino energies but otherwise compute the moments numerically with the scattering kernel. 

In Fig.~\ref{fig:critical_domains_Sigma_0} we summarize the domains in the $\The-x$ plane for which one can expect very efficient isotropization (upper panel) and neutrino redistribution in energy or energy exchange (lower panel) by the scattering with electrons and positrons. The first is determined by asking when $\tau^*_{\rm sc}=1$, while the latter follows from $|y^*_{\rm sc}|=1$. For comparison, we also given the domain boundaries that follow from using the simpler moment expressions in the Doppler- or recoil-dominated regimes \citep[see section~2.1.1 and 2.1.2 of][]{Chluba2026nuKernels}.

For the isotropization, the leading order scaling of the total cross section with neutrino energy implies that at low energies, isotropization becomes inefficient significantly earlier than at high energies. We see that at $x\ll 1$ the exact kernel domain approaches the domain boundary obtained using the Doppler dominated limit. However, given that the temperature is quite high around the boundary, the Doppler limit is reached rather slowly. In contrast, the high energy asymptote is approached closely within the considered range of neutrino energies. Importantly, around the maximum of the neutrino Fermi-Dirac distribution, neutrinos essentially freely stream with respect to $\nu_e-e^\pm$ scattering during the annihilation era.

A similar picture can be drawn for the efficiency of energy exchange with the electrons and positrons from the lower panel of Fig.~\ref{fig:critical_domains_Sigma_0}. Here, we have two distinct regimes, where neutrinos primarily up-scatter at $x\lesssim 4$ through Doppler boosting (i.e., $y^*_{\rm sc}>0$), while neutrinos down-scatter at $x\gtrsim 4$ primarily through recoil (i.e., $y^*_{\rm sc}<0$). Neutrinos injected at around $x\simeq 4$ will not drift much in energy by interactions with the $e^\pm$-pairs but primarily broaden as is also known from Compton scattering \citep{Chluba2015GreensII}. In addition, we can conclude that during the annihilation era, at $x\lesssim 100$ most neutrinos not only free stream (i.e., do not change their propagation direction) but also do not exchange much energy with the $e^\pm$-plasma. To have significant energy exchange, we have to go to phases with $\The \gtrsim 2-5$. We also remark that this conclusion should not change much for $\nu_{\tau/\mu}-e^\pm$ scattering.

We also mention that at $x\lesssim 1$, the transition to efficient energy exchange with the $e^\pm$ pairs lies at lower temperatures than the one for isotropization. This implies that while there are few scatterings, the fractional energy exchange per scattering exceeds unity. This is in stark contrast to the CMB, for which energy redistribution becomes inefficient at redshift $z\lesssim 50,000$, while isotropization becomes inefficient around last scattering at $z\lesssim 10^3$ \citep{Hu1995PhD}.

Finally, in Fig.~\ref{fig:critical_domains_Sigma_0}, we also illustrate the effects of final state blocking on the critical temperature. The lines close to the 'full kernel' domain boundaries include the effect of final state electron blocking (long dashed) and final state electron and neutrino blocking (short dashed). The effects are mainly visible at low neutrino energies, when the factors $(1-f_3)(1-f_4)\rightarrow 1/4$. However, the overall conclusions are not affected much.

%------------------------------------
\section{Conclusions}
\label{sec:conclusions}
%------------------------------------
In this work we studied the evolution of neutrino distributions under repeated elastic scattering with thermal electrons under various conditions. Starting from the redistribution kernel formalism, we formulated the collision problem in a form that retains recoil, relativistic kinematics and Fermi-blocking while remaining suitable for efficient numerical evaluation. To break the problem down, we then also derived the corresponding Fokker--Planck (FP) limit of the collision term in two independent ways, using the energy-transfer moments and the boost-operator approach. We then discussed analytic solutions to the related FP equations in limiting cases to provide useful intuition about the dynamics of scattering problems. With the boost operator approach, we were also able to obtain the Doppler-dominated scattering operator to all orders in the electron momentum [see Eq.~\eqref{eq:S_SZ_nu}], which we used to construct the relativistic single-scattering, or $\nu$SZ, expansion [see Eq.~\eqref{eq:Dn_nu}]. 

The comparison with the photon problem proved useful throughout and helped us identify which aspects of the redistribution are generic and which follow from the different interaction physics and quantum statistics. We implemented numerical treatments of both the FP equation and the full scattering kernel. Single line-injection calculations illustrate that the FP description captures the overall drift and broadening well when the energy transfer per scattering is small, while the full kernel becomes important for narrow features and hotter scattering media. In particular, the exact treatment retains the non-local redistribution and explicit temperature dependence that are not captured by the local diffusion approximation (see Fig.~\ref{fig:line_evo_Kernel_I} and \ref{fig:line_evo_Kernel_III}). For sufficiently long evolution, the solutions approach the expected number-conserving kinetic-equilibrium distributions. 
In addition we illustrated that for thermalization problems involving small background distortions the obtained spectral shapes are all very similar (see Fig.~\ref{fig:thermal_evo_Kernel_I} and \ref{fig:thermal_evo_Kernel_II}), even for varying background temperatures. The main difference stems from the temperature dependence of the overall scattering efficiency, which leads to differences in the timing of the distortions rather than the shape. This may provide useful avenues for simplifying more detailed computations, where an effective FP description could be found by suitable coordinate transformations. However, we leave this exploration to future work.

As a simple cosmological application, we estimated the efficiency of neutrino--electron and neutrino--positron scattering around the epoch of electron--positron annihilation. The zeroth and first moments of the kernel allow us to distinguish between efficient isotropization and efficient energy exchange domains. Around the main thermal part of the neutrino distribution, both become inefficient with respect to $e^\pm$ scattering during the annihilation era, while significant energy exchange occurs only at earlier, hotter stages (see Fig.~\ref{fig:critical_domains_Sigma_0} for details). This provides a simple illustration of how the kernel moments can be used to identify the relevant transport regimes. 

The methods developed here should be useful in more detailed neutrino transport problems. In the early Universe, one could combine the present treatment with the other collision processes relevant for neutrino decoupling and study the evolution of non-thermal spectral features caused by energy injection problems. In supernovae and other dense environments, the same approach can be extended to scattering with nucleons and used as a benchmark for approximate transport schemes. It will also be interesting to generalize the present isotropic treatment to include bulk motions and anisotropic scattering, either directly at the level of the redistribution kernels \citep[e.g.,][]{Lee2024SZpack} or using the boost-operator formalism \citep{Hoey2026kSZ}. 
Finally, we have also not included flavor-oscillations or looked at a wider range of cases where degeneracy effects may become more significant.
We leave these extensions to future work.

\section*{Acknowledgments}
{\small This work was supported by the UKSA grant: LiteBIRD UK ST/Y005945/1. BC was supported by an NSERC Banting Fellowship, as well as the Simons Foundation (Grant Number 929255). M.U. is supported by IBS under the project code, IBS-R018-D3. M.Y. is supported by IBS under the project code, IBS-R018-D3, and by JSPS Grant-in-Aid for Scientific Research Number JP23K20843.
This work was also supported in part by a grant of access to OpenAI models through the ChatGPT for Academic Researchers program. Feedback on the manuscript and checks of some of the derivations were performed using ChatGPT. In addition, some of the interface development with previously developed codes and figure setups were done with the help of Codex. However, the authors take full responsibility for the presentation and correctness of the results.}

{\small
\bibliographystyle{plain}
\bibliography{bibliography,Lit-all}
}

\end{document}

%% file: preamble.tex
\usepackage{graphicx}
\usepackage{float}
\usepackage{hyperref}
\usepackage{xcolor, xstring}
\usepackage{amssymb, amsmath, mathtools}
\usepackage{listings}
\usepackage{lscape}
\usepackage{lipsum}
\usepackage{multicol}
\usepackage{physics}
\usepackage{natbib}
\usepackage{txfonts}
\usepackage[small]{caption}
\usepackage{xspace}
\usepackage{comment}

\newcommand{\pot}[2]{#1 \times 10^{#2}}

\newcommand{\boostO}{\hat{\mathcal{O}}_x}

\newcommand{\DiffO}{\hat{\mathcal{D}}_x}

\newcommand{\nbb}{n_{\rm bb}}
\newcommand{\Gspec}{{G}}
\newcommand{\Yspec}{{{Y}}}
\newcommand{\Ynspec}[1]{{Y}_{#1}}

%% file: Befehle.tex
\newcommand{\expf}[1]{{{\rm e}^{#1}}}

\newcommand{\cm}{{\rm cm}}

\newcommand{\vgh}{{\hat{\boldsymbol\gamma}}}

\newcommand{\id}{{\,\rm d}}

\newcommand{\beq}{\begin{equation}}   %

\newcommand{\eeq}{\end{equation}}   %

\newcommand{\beqa}{\begin{eqnarray}}   %

\newcommand{\eeqa}{\end{eqnarray}}   %

\newcommand{\bealf}[1]{\begin{align} #1 \end{align}}

\newcommand{\beal}{\begin{align}}
\newcommand{\enal}{\end{align}}

\newcommand{\bspl}{\begin{split}}

\newcommand{\espl}{\end{split}}

\newcommand{\bsub}{\begin{subequations}}

\newcommand{\esub}{\end{subequations}}

\newcommand{\bmulti}{\begin{multline}}   %

\newcommand{\beqm}{\begin{mathletters}}   %

\newcommand{\eeqm}{\end{mathletters}}   %

\newcommand{\me}{m_{\rm e}}

\newcommand{\Ne}{N_{\rm e}}

\newcommand{\Te}{T_{\rm e}}

\newcommand{\Tg}{T_{\gamma}}

\newcommand{\The}{\theta_{\rm e}}

\newcommand{\Thg}{\theta_{\gamma}}

\newcommand{\sigT}{\sigma_{\rm T}}